\documentclass[11pt,a4paper]{article}
\usepackage{float}
\usepackage{jheppub}
\usepackage{amsmath,amssymb,mathrsfs,bm,graphicx,booktabs,microtype,placeins,needspace,array}
\usepackage{siunitx}
\AtBeginEnvironment{figure}{\setlength{\abovecaptionskip}{6pt}}
\apptocmd{\thebibliography}{\setlength{\itemsep}{1pt plus 0.2ex minus 0.05ex}}{}{}
\let\fulltableofcontents\tableofcontents
\renewcommand{\tableofcontents}{\begingroup\small\fulltableofcontents\endgroup}
\renewcommand{\afterTocRuleSpace}{\clearpage}
\newcommand{\dd}{\mathrm{d}}
\newcommand{\ii}{\mathrm{i}}
\newcommand{\rh}{r_{\mathrm h}}
\newcommand{\cW}{\mathcal W}
\newcommand{\cV}{\mathcal V}
\newcommand{\Om}{\Omega}
\newcommand{\Log}{\operatorname{Log}}

\title{Metric and dilaton information from high-overtone quasinormal modes of analytic Einstein--Maxwell--dilaton black holes}
\hypersetup{pdftitle={Metric and dilaton information from high-overtone quasinormal modes of analytic Einstein--Maxwell--dilaton black holes},pdfauthor={Yu Shi, Chikun Ding, Yuebing Zhou, Changqing Liu, Weike Deng, Sheng Long}}
\author[a,*]{Yu Shi,}
\author[a]{Chikun Ding,}
\author[a]{Yuebing Zhou,}
\author[a]{Changqing Liu,}
\author[b]{Weike~Deng,}
\author[c,*]{and~Sheng~Long\note[*]{Corresponding authors}}
\affiliation[a]{Department of Physics, Huaihua University, Huaihua, Hunan 418008, China}
\affiliation[b]{School of Science, Hunan Institute of Technology, Hengyang 421002, China}
\affiliation[c]{School of Fundamental Physics and Mathematical Sciences, Hangzhou Institute for Advanced Study, University of Chinese Academy of Sciences, Hangzhou 310024, China}
\emailAdd{shiyu@hhtc.edu.cn}
\emailAdd{shenglong@ucas.ac.cn}
\abstract{Black-hole interiors are hidden from exterior observers, yet their geometry can enter the high-overtone spectrum of linear perturbations. We study how this spectrum encodes metric and supporting-matter information in analytic Einstein--Maxwell--dilaton black holes. Coupling a neutral spectator to the dilaton at fixed UV dimension provides a matter-sensitive channel alongside the momentum response. For real-endpoint spacelike connections, the momentum response and mass-label dependence separate the metric exponents from the signed dilaton slope. Near the causal threshold, a complex endpoint controls the spectral family: the mass profiles approach a common value, producing a shared fractional correction, while regular propagation retains label dependence. This mechanism is present at criticality without an inner horizon and continues locally into timelike interiors. Exterior numerical spectra support the reconstruction and the leading complex-endpoint responses. Local scaling and probe coupling set the candidate responses, while wave connection determines their imprint on the exterior spectrum.}

\begin{document}
\maketitle
\raggedbottom

\section{Introduction}\label{sec:intro}
In classical general relativity, future-directed causal signals from inside an event horizon cannot reach the asymptotic region relative to which that horizon is defined~\cite{Wald2001Thermodynamics}. Quantities defined in the exterior or at the boundary can still encode features of the geometry behind the horizon. Determining which features they reveal is an inverse problem: it concerns the causal character of the singularity, contraction and stretching along different directions, and the behavior of the matter supporting the geometry. Where a spacelike region admits a local Kasner regime, power-law exponents describe its anisotropic approach to the singularity~\cite{FrenkelEtAl}. More general evolution can pass through successive regimes~\cite{BKL1970,DamourHenneauxNicolai2003}. Two questions motivate our analysis: how an exterior spectrum can distinguish the supporting matter from the metric, and why local interior scaling need not appear in that spectrum.

The supporting matter can shape both the local scaling and the causal structure. Charged scalar condensates can induce Kasner inversions~\cite{HartnollSuperconductor,CaiDuanLiYang2024}, while oscillatory Kasner transitions occur in vector-supported holographic black-hole interiors~\cite{CaiGeLiYang2022,DeClerckHartnollSantos2024,CaiDuanLiYang2025}. In holographic scalar models, the matter profile can replace a would-be inner horizon with a spacelike Kasner region~\cite{HartnollInterior}. In spontaneously scalarized asymptotically flat black holes, amplification of a small exterior scalar near the would-be Cauchy horizon generates critical scaling of the terminal Kasner parameter~\cite{SunYangCaiLiYang2026}. For spherical and planar black holes with charged scalar hair, conserved-charge arguments exclude a smooth inner Cauchy horizon, including extensions to nonlinear electrodynamics~\cite{CaiNoInnerHorizon,AnLiYang2021}. Other charged stationary matter systems can retain an inner horizon followed by a timelike singularity~\cite{AreanEtAl}.

Thermal observables reveal interior geometry through several complementary signatures. Analytic continuations of large-dimension two-point functions encode singularity data in complex geodesic or WKB saddles~\cite{FidkowskiEtAl2004,FestucciaLiu}, including the shock-induced bending of the Schwarzschild--AdS singularity~\cite{HorowitzLeungQueimadaZhao2024}. Boundary operator product expansions (OPEs) capture related structure: the multi-stress-tensor sector reproduces bouncing-geodesic signatures in AdS black branes~\cite{CeplakEtAl2024}, while the stress-tensor and current sector gives evidence for timelike-singularity signatures beyond a charged black brane's inner horizon~\cite{CeplakLiuParnachevValach2026}. Sensitivity depends on the available saddles: a self-dual axion black hole has a curvature singularity but no bouncing geodesic attached to its conformal boundary~\cite{GrozdanovValachVrbica2026}.

Other observables resolve proper time or radial scaling. Large-mass analytic continuation of thermal one-point functions of curvature-coupled fields can yield the proper time from the horizon to the singularity~\cite{GrinbergMaldacena2021}; higher-point correlators of interacting heavy fields contain related WKB phases~\cite{Singhi2025ProperTime}. Thermal $a$-functions follow radial evolution beyond the horizon~\cite{CaceresKunduPatraShashi2022,CaceresShashi2023Anisotropic}, with a near-singularity power fixed by the local scaling geometry in the analytic EMD family~\cite{AreanEtAl}.

Geometric probes reveal further combinations of interior data. Holographic complexity uses bulk volumes~\cite{StanfordSusskind2014} or Wheeler--DeWitt actions~\cite{BrownEtAl2016ComplexityAction}; the spacelike singularity contributes explicitly to the latter's growth in single-horizon charged dilaton black holes~\cite{CaiSasakiWang2017}. The prescription matters: at the superconducting transition in an M-theory model, late-time complexity=action growth jumps while complexity=volume growth remains continuous~\cite{AnLiYangYang2022}. Generalized complexity~\cite{BelinEtAl2022ComplexityAnything} includes constant-mean-curvature constructions that, at large mean curvature, approach the AdS Schwarzschild singularity and measure local volume and curvature~\cite{JorstadMyersRuan2023}.

Entanglement-based observables also sample interior geometry. Linear growth of spatial-subregion entanglement in thermofield-double and related pure states reflects the stretching of extremal surfaces along an interior spacelike direction~\cite{HartmanMaldacena2013}. Timelike entanglement extends these constructions by analytic continuation to generally complex quantities, with a pseudoentropy interpretation and candidate holographic prescriptions~\cite{Doi2023Timelike}. Such prescriptions can detect curvature singularities~\cite{Anegawa2024Timelike} and different causal arrangements of spacelike singularities in charged scalar-hairy models~\cite{Li2026Timelike}. In the piecewise extremal-surface prescription, Kasner asymptotics and the null energy condition ensure an interior critical surface governing the real part's late-time linear growth in single-horizon AdS black holes~\cite{LiYang2026TimelikeGrowth}. These examples connect interior sensitivity to how the observable samples the geometry. Linear waves provide a spectral approach in which that relation can be followed through the wave equation and its exterior boundary conditions.

\Needspace{3\baselineskip}
Quasinormal modes (QNMs) are resonances of linear perturbations about a black-hole background. Their complex frequencies follow from the perturbation equation and physical boundary conditions, and characterize oscillation and decay in linear response~\cite{BertiCardosoStarinets2009}. For our AdS scalar spectator, the conditions are ingoing behavior at the future outer horizon and a vanishing source at the AdS boundary~\cite{HorowitzHubeny}. With standard quantization and the holographic retarded prescription, these frequencies are poles of the retarded two-point function of the boundary operator dual to the spectator~\cite{SonStarinets,KovtunStarinets2005,HartnollKumar2005}. We focus on high overtones, where analytic continuation gives access to the relation between the exterior spectrum and interior scaling.

High-damping quantization connects this exterior problem to the effective potential near a singularity through continuation in complex radius. Monodromy and complex-WKB analyses established this relation for asymptotically flat black holes~\cite{MotlNeitzke,AnderssonHowls2004}; extensions to de Sitter and AdS incorporated the corresponding outer boundary conditions~\cite{CardosoNatarioSchiappa,NatarioSchiappa2004}. For spherical power-law singularities, the leading scalar-wave Bessel index depends on a combination of the angular metric power and the sphere dimension~\cite{DasShankaranarayanan2005,GhoshShankaranarayananDas2006}. The leading spectrum therefore probes a particular combination of local data; recovering individual exponents calls for additional spectral information. First-order perturbations about the near-singularity Bessel equation already yielded noninteger-power corrections for four-dimensional Schwarzschild scalar and gravitational waves~\cite{MusiriSiopsisSchwarzschild2003} and Schwarzschild--AdS gravitational perturbations~\cite{MusiriNessSiopsis2006}.

The asymptotic spectrum combines local scaling with global propagation. With appropriate analyticity and growth properties, holographic thermal two-sided correlators admit product representations over QNM poles, relating high-frequency spectral data to the OPE and complex-time singularities~\cite{DodelsonIossaKarlssonZhiboedov2024}. For mixed time--momentum retarded correlators, overlapping OPE and QNM convergence domains relate high-overtone asymptotics to OPE singularities~\cite{ArnaudoIossaKarlssonWithers2026}. At fixed probe parameters, AdS quantization involves optical propagation and boundary matching as well as near-singularity indices~\cite{NatarioSchiappa2004,GhoshShankaranarayananDas2006,MusiriNessSiopsis2006}. Local powers can determine exponents of algebraic corrections, while their coefficients and the overall spacing retain information about the background and the connection. In Schwarzschild spacetime, complex-time bouncing singularities also set the convergence region of the QNM expansion of the retarded Green function~\cite{ArnaudoWithers2026}. The frequency limit matters: in a nonsingular quantum-corrected single-horizon model, the zero-correction and infinite-damping limits do not commute~\cite{BabbDaghighKunstatter2011}. Bohr--Sommerfeld formulas organize related spectral data in the large-mass or large-momentum limits~\cite{FestucciaLiuWKB}. In a reflecting-cavity construction, the asymptotic spacing is related to the analytically continued null travel time from the wall to the singularity and back~\cite{GrozdanovMovrinValach,GrozdanovMovrinValachCavity2026}.

Different spectral responses now allow interior parameters to be distinguished. Momentum-dependent high-overtone spectra resolve Kasner-sensitive nonanalytic scaling and transitions between distinct Kasner regimes in scalar-hairy examples. A probe-mass response provides another route to the temporal metric exponent~\cite{HartnollZhiboedov}. Distinct spectral structures allow independent recovery of temporal and spatial metric exponents without imposing Kasner constraints, as demonstrated in five-dimensional Schwarzschild and planar Schwarzschild--AdS backgrounds and in the charged dilatonic Gibbons--Maeda family~\cite{XiaoNianLi}. In quantum BTZ geometries, spectral offsets and angular-momentum corrections relate the spectrum to singular scaling and exhibit overtone crossovers between interior regimes~\cite{CovielloEtAl,CovielloEtAlLong}.

Two aspects of these developments guide the present work: adding direct sensitivity to the supporting matter, and understanding how wave propagation selects the local information visible in the spectrum. An analytic family of planar Einstein--Maxwell--dilaton (EMD) black holes~\cite{HendiSheykhiDehghani2010,GouterauxKiritsis2011}, with an established parameterization and interior classification~\cite{AreanEtAl}, provides a setting in which both can be studied. Its closed-form metric and matter fields supply a geometric reference for spectral inference. Varying the dilaton coupling changes the theory and takes the interior from a spacelike singularity to a timelike singularity behind an inner horizon; the critical geometry remains spacelike. Changing the charge at fixed coupling instead compares states of the same theory and changes the global optical propagation.

Momentum and ordinary mass responses distinguish temporal and spatial metric scaling~\cite{HartnollZhiboedov,XiaoNianLi}. To resolve the signed dilaton slope directly, we couple a neutral spectator to the dilaton through a tunable exponential profile. Changing its label varies the matter weight, while subtracting its boundary value preserves the spectator's UV dimension. Momentum and two distinct dilaton-coupling labels then provide separate responses within the same wave problem. Their powers reveal how geometry and supporting matter enter the local potential; the wave connection to the UV boundary and outer horizon determines which of these contributions enter the QNM poles.

For spacelike real-endpoint connections, the mass-label intercept and slope separate temporal metric and signed dilaton scaling; momentum supplies the transverse metric information. At criticality, a complex endpoint controls the spectral family considered here. The mass profiles share a limiting value there and hence a fractional mass correction, while regular propagation retains label dependence. This mechanism continues locally across the causal threshold. The timelike examples relate propagation and endpoint contributions to radial interior scaling. Together, probe coupling and wave connection determine the spectral imprint of interior scaling.

\section{Setup}\label{sec:geometry}
\subsection{Analytic EMD geometry and the causal threshold}
The background is a family of charged, planar, asymptotically AdS black holes supported by a neutral dilaton $\phi$. Its closed-form fields let us compare the information extracted from waves with the interior geometry. Set the AdS radius and $16\pi G$ to one. For boundary spacetime dimension $d>2$, the canonically normalized action of Ref.~\cite{GouterauxKiritsis2011} reads
\begin{equation}\label{eq:action}
 I_{\rm EMD}=\int\dd^{d+1}x\sqrt{-g}
 \left[R-\frac12(\partial\phi)^2-\frac14e^{-(d-2)\delta\phi}F^2
       +\cV_\delta(\phi)\right].
\end{equation}
Here $F=\dd A$ and $\delta>0$ labels the theory. The three-exponential potential and charged planar solution are given in Ref.~\cite{GouterauxKiritsis2011}.\footnote{The potential and solution appear in Eqs.~(8.2) and (8.9)--(8.12), with their $p=d$ and $V_0=d(d-1)$. The planar solution is equivalent to the $k=0$ geometry in Eqs.~(6)--(13) of Ref.~\cite{HendiSheykhiDehghani2010} after canonical normalization of the dilaton and Maxwell fields and the radial change $r^{d-2}=\varrho^{d-2}-b^{d-2}$, with $Q=b^{d-2}$ and $M=c^{d-2}$; $\varrho$ is their original radial coordinate.} We use the background parameterization and regular Maxwell field of Ref.~\cite{AreanEtAl}; in our sign convention $\cV_\delta(0)=d(d-1)$ and $m_\phi^2=-2(d-2)$. The metric and dilaton are
\begin{equation}\label{eq:background}
 \dd s^2=H^{\beta_g}\left[-fH^{-\gamma}\dd t^2+
               \frac{\dd r^2}{f}+r^2\dd\vec x_{d-1}^{\,2}\right],
\end{equation}
\begin{equation}\label{eq:backgrounddilaton}
 \phi=-\frac{2(d-1)\delta}{2+(d-2)(d-1)\delta^2}\log H.
\end{equation}
\begin{equation}\label{eq:radialfunctions}
 H(r)=1+\frac{Q}{r^{d-2}},\qquad
 f(r)=r^2H^\gamma-\frac{M}{r^{d-2}}.
\end{equation}
The metric powers are
\begin{equation}\label{eq:metricpowers}
 \beta_g=\frac{4}{(d-2)[2+(d-2)(d-1)\delta^2]},
 \qquad \gamma=(d-1)\beta_g.
\end{equation}
A nonextremal outer horizon $\rh$ fixes $M=\rh^dH(\rh)^\gamma$. The parameter $Q>0$ controls the charge and dilaton profile, but is distinct from the conserved Maxwell flux. Thus $Q/\rh^{d-2}$ varies a state at fixed action; varying $\delta$ changes the theory.

The causal classification follows directly from the blackening function. Write $\cW=r^dH^\gamma$, so $f=(\cW-M)/r^{d-2}$. The power of $\cW$ at $r=0$ changes sign when $d-(d-2)\gamma=0$, giving the critical coupling~\cite{AreanEtAl}
\begin{equation}\label{eq:causalthreshold}
 \delta_c^2=\frac{2}{d(d-1)}.
\end{equation}
Above the threshold, $\cW$ grows monotonically from zero and $f<0$ throughout the interior: the singularity is spacelike. At the threshold the two terms of $f$ have the same power and combine as $f\sim-(M-Q^\gamma)r^{-(d-2)}$; since $M>Q^\gamma$, the singularity remains spacelike. Below it, $\cW$ diverges at the origin and has one minimum. An inner horizon then separates the $f<0$ interior from the $f>0$ region ending at a timelike singularity. The designated outer horizon satisfies
\begin{equation}\label{eq:outerhorizondomain}
 \rh^{d-2}>\left[\frac{(d-2)\gamma}{d}-1\right]Q .
\end{equation}
Equality is the extremal limit. The causal threshold is distinct from extremality. We study these stationary geometries, whose Cauchy horizons can be dynamically unstable~\cite{PoissonIsrael1990}. Under the relevant energy and horizon-geometry conditions, horizon-counting arguments in static and stationary-axisymmetric geometries allow at most one nondegenerate inner Killing horizon per connected interior branch~\cite{YangCaiLi2022,YeLiYangLi2026}.

For $d=3$, used below, the critical coupling $\delta^2=1/3$ gives the Gubser--Rocha model, whose black-hole solution admits an eleven-dimensional uplift asymptotic to $\mathrm{AdS}_4\times S^7$~\cite{AreanEtAl,GubserRocha2010}.

At a spacelike tip, proper time gives
\begin{equation}\label{eq:kasner}
\dd s^2\simeq-\dd\tau^2+c_t\tau^{2p_t}\dd t^2
                  +c_x\tau^{2p_x}\dd\vec x_{d-1}^{\,2},\qquad
 \phi\simeq-\sqrt2p_\phi\log\tau+\phi_0.
\end{equation}
The coordinate $t$ is temporal outside the horizon and labels a spatial direction in this limit. The frequency analysis below uses $d=3$, so we write $a=\delta^2$ and $a_c=1/3$. The exact metric and supporting dilaton give~\cite{AreanEtAl}\footnote{For general $d$, the same solution gives
\[
p_t=\frac{2-(d-2)(d-1)\delta^2}{2+d(d-1)\delta^2},\quad
 p_x=\frac{2(d-1)\delta^2}{2+d(d-1)\delta^2},\quad
 p_\phi=-\frac{2\sqrt2(d-1)\delta}{2+d(d-1)\delta^2}.
\]
They obey $p_t+(d-1)p_x=1$ and $p_t^2+(d-1)p_x^2+p_\phi^2=1$.}
\begin{equation}\label{eq:exactp}
 (p_t,p_x,p_\phi)=\frac{(1-a,\ 2a,\ -2\sqrt{2a})}{1+3a}.
\end{equation}
The background constraints are
\begin{equation}\label{eq:kasnerconstraints}
 p_t+2p_x=1,\qquad p_t^2+2p_x^2+p_\phi^2=1.
\end{equation}
The dilaton contribution to the quadratic relation permits all-positive spatial powers, a familiar possibility in massless-scalar near-singularity solutions~\cite{BelinskiiKhalatnikov1973}. For $1/3\le a<1$, both metric exponents are positive and all spatial directions contract as $\tau\to0$. At $a=1$, $p_t$ vanishes: the leading proper length along the interior $t$ direction tends to a constant. For $a>1$, that direction stretches while the transverse directions continue to contract. The spatial volume still vanishes as $\tau^{p_t+2p_x}=\tau$. As $a\to\infty$, the triplet approaches the vacuum Kasner values $(-1/3,2/3,0)$, and the dilaton contribution to the quadratic Kasner relation tends to zero. The change at $a=1$ will also organize the momentum responses in section~\ref{sec:spacelikepropagation}.

\begin{figure}[!htbp]
\centering\includegraphics[width=\linewidth,trim=0 10.4bp 0 0,clip]{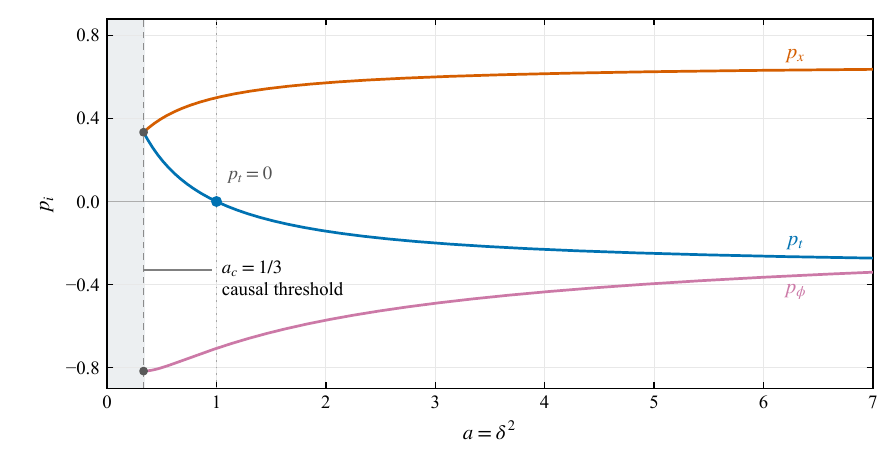}
\caption{Interior metric and dilaton scaling in the analytic EMD family, $d=3$. Solid curves give the analytic indices~\eqref{eq:exactp} as functions of $a=\delta^2$. The spacelike branch includes $a_c=1/3$; the shaded interval $0<a<a_c$ has the timelike proper-radial scaling~\eqref{eq:timelikepowers}. The marker at $a=1$, where $p_t=0$, separates contraction and stretching of the interior $t$ direction.}\label{fig:geometry}
\end{figure}

At the causal threshold the triplet is $(1/3,1/3,-\sqrt{2/3})$ and all spatial directions contract equally. For $0<a<1/3$, proper radial distance instead gives
\begin{equation}\label{eq:timelikepowers}
\dd s^2\simeq\dd\sigma^2+\sigma^{2a}
                  (-c_t\dd t^2+c_x\dd\vec x_2^{\,2}),\qquad
 \phi\simeq2\sqrt a\log\sigma+\phi_0.
\end{equation}
This is radial scaling towards a timelike singularity, rather than a continuation of the proper-time Kasner triplet. Figure~\ref{fig:geometry} shows how the interior scaling of the analytic background varies with the dilaton coupling, providing a geometric reference for the spectral analysis below.

The supporting dilaton and the spectator have different boundary conditions. In the area coordinate $z_A=g_{xx}^{-1/2}$, the exact dilaton obeys a mixed scalar boundary condition~\cite{Witten2001Multitrace}. Appendix B of Ref.~\cite{AreanEtAl} gives the model-specific relation $\phi_{(1)}=-(1-a)\phi_{(0)}^2/(4\delta)$, where $\phi=z_A\phi_{(0)}+z_A^2\phi_{(1)}+\cdots$. The spectator introduced next obeys standard quantization with vanishing source.

\subsection{Fixed-dimension scalar-dependent probes}
Let $\chi$ be a neutral spectator, distinct from $\phi$. Starting from the standard quadratic scalar action with a canonical kinetic term and constant mass~\cite{SonStarinets}, we introduce adjustable dilaton-dependent mass profiles:
\begin{equation}\label{eq:probeaction}
I_\chi=-\frac12\int\dd^{d+1}x\sqrt{-g}
 \left[(\partial\chi)^2+
 \left(m_\chi^2+\sum_\ell\eta_\ell W_\ell(\phi)\right)\chi^2\right],
\end{equation}
Mass and momentum responses can isolate singularity-sensitive terms in QNM spectra~\cite{HartnollZhiboedov}. Here we choose the mass profile
\begin{equation}\label{eq:massprofile}
 W_\ell(\phi)=e^{-\ell\phi}-1,\qquad W_\ell(0)=0.
\end{equation}
The exponential converts the dilaton's logarithmic interior behavior into a label-dependent power. Subtracting one makes the profile vanish at the AdS boundary, so varying $\eta_\ell$ leaves the UV mass and $\Delta_+=d/2+\sqrt{d^2/4+m_\chi^2}$ unchanged. We set $m_\chi=0$ in the spectral calculations. Two distinct labels $\ell$ thus couple the same metric weight to different dilaton powers, separating the matter contribution in the spectral responses.

For the critical comparison we introduce an auxiliary spatial-gradient insertion to compare different dilaton weights, with $i,j$ running over the black brane's planar spatial directions:
\begin{equation}\label{eq:gradientaction}
\Delta I_g=-\frac{\eta_g}{2}\int\dd^{d+1}x\sqrt{-g}\,
 G_{\ell_g}(\phi)g^{ij}\partial_i\chi\partial_j\chi,
\end{equation}
Here the profile and independent wave parameter are
\begin{equation}\label{eq:gradientprofile}
 G_{\ell_g}(\phi)=e^{-\ell_g\phi},\qquad \zeta=\eta_gk^2.
\end{equation}
The measured response is
\begin{equation}\label{eq:mixedgradientresponse}
 R_g(n;\ell_g)=
 \left.\partial_{\eta_g}\partial_{k^2}\omega_n\right|_0
 =\left.\partial_\zeta\omega_n\right|_0.
\end{equation}
The first $\eta_g$ derivative at $k=0$ vanishes; the mixed derivative inserts $G_{\ell_g}$ into the momentum term. Its limit $G_{\ell_g}(0)=1$ changes the spatial kinetic coefficient but preserves the UV indices.

For the $d=3$ background, the same profiles can be written as
\begin{equation}\label{eq:profilepowers}
 b_\ell=\frac{2\ell\sqrt a}{1+a},\qquad
 W_\ell=H^{b_\ell}-1,\qquad G_{\ell_g}=H^{b_{\ell_g}}.
\end{equation}
The label is fixed when differentiating a probe strength; $b_\ell$ also depends on the background.

For $\chi=e^{-\ii\omega t+\ii\vec k\cdot\vec x}\mathcal U(r)$, the determinant and radial inverse metric give $\sqrt{-g}=r^{d-1}H^{\beta_g}$ and $\sqrt{-g}g^{rr}=r^{d-1}f$. The wave equation is
\begin{equation}\label{eq:radial}
 \frac1{r^{d-1}}\partial_r\!\left(r^{d-1}f\partial_r\mathcal U\right)
 +\left[\frac{\omega^2H^\gamma}{f}
 -\frac{k^2+\zeta G_{\ell_g}}{r^2}
 -H^{\beta_g}\left(m_\chi^2+\sum_\ell\eta_\ell W_\ell\right)\right]
 \mathcal U=0 .
\end{equation}
Define the metric weights and Liouville transformation on the exterior branch by
\begin{equation}\label{eq:liouvillemap}
\begin{gathered}
 \mathsf D=-g_{tt}=fH^{\beta_g-\gamma},\qquad
 \mathsf C=g_{xx}=r^2H^{\beta_g},\\[3pt]
 \frac{\dd r_*}{\dd r}=\frac{H^{\gamma/2}}{f},\qquad
 \Psi=S\mathcal U,\qquad S=\mathsf C^{(d-1)/4}.
\end{gathered}
\end{equation}
The continued Schr\"odinger equation and its potential are
\begin{equation}\label{eq:mainpotential}
\begin{gathered}
 \frac{\dd^2\Psi}{\dd r_*^2}+\bigl[\omega^2-V(r)\bigr]\Psi=0,\\[4pt]
 V=V_g+m_\chi^2\mathsf D+k^2\frac{\mathsf D}{\mathsf C}
       +\sum_\ell\eta_\ell\mathsf DW_\ell
       +\zeta\frac{\mathsf DG_{\ell_g}}{\mathsf C},\qquad V_g=\frac{S''}{S}.
\end{gathered}
\end{equation}
Primes denote optical derivatives. The momentum term weights the geometry through $\mathsf D/\mathsf C$, while the mass profile enters through $\mathsf D W_\ell$. The gradient insertion adds the dilaton weight $G_{\ell_g}$ to the momentum factor. These weights make the three spectral responses sensitive to different combinations of metric and dilaton scaling.

\subsection{Exterior poles and response derivatives}\label{sec:probe}
The retarded problem imposes $\mathcal U_{\rm in}\sim e^{-\ii\omega r_*}$ at the nonextremal outer horizon and the normalizable $\mathcal U_{\rm UV}\sim r^{-3}$ branch at the AdS boundary. Normalize each leading endpoint coefficient to one. If $\mathcal E$ is the UV source coefficient of the ingoing solution, its simple zeros are QNMs~\cite{HorowitzHubeny,SonStarinets}. Analytic continuation is a way to compute this same exterior source, with no additional boundary condition at the real or complex singularities.

Differentiating $\mathcal E(\omega_n(\xi),\xi)=0$ gives the simple-root response formula, as in the Wronskian construction of appendix D, Eqs.~(D.8)--(D.9), of Ref.~\cite{HartnollZhiboedov}:
\begin{equation}\label{eq:evansidentity}
\begin{aligned}
 \mathcal E(\omega_n,0)&=0,\qquad
 \mathcal E_\omega(\omega_n,0)\ne0,\\
 R_\xi(n)&=\left.\partial_\xi\omega_n\right|_0
       =-\left.\frac{\mathcal E_\xi}{\mathcal E_\omega}
                    \right|_{(\omega_n,0)}.
\end{aligned}
\end{equation}
The directions $\xi=k^2,\eta_\ell,\zeta$ give $R_k,R_\ell,R_g$, respectively. All derivatives are evaluated at zero momentum and probe strengths, at fixed background and specified labels. They are independent parameter directions of the same spectator problem. Any analytic nonzero rescaling of $\mathcal E$ preserves these quotients at a root. Appendix~\ref{app:source} gives the source variation, including the endpoint terms needed when a profile changes subleading UV propagation.

\section{Spacelike interiors: metric and dilaton exponents}\label{sec:connection}
\subsection{Local exponents and dilaton sensitivity}
The spacelike analysis separates two steps: how the metric and dilaton enter the local wave potential, and how the wave connection carries those contributions into the exterior poles. Near the real spacelike tip, the linear Kasner relation fixes the optical scaling
\begin{equation}\label{eq:spaceliketipscaling}
 s\propto\tau^{1-p_t},\qquad S\propto s^{1/2},\qquad
 V_g\sim-\frac1{4s^2}.
\end{equation}
In generic power-law singularity analyses, the leading Bessel index depends on the angular metric power and sphere dimension~\cite{DasShankaranarayanan2005,GhoshShankaranarayananDas2006}. Here the linear Kasner relation fixes the same index for different metric--dilaton triplets. Their distinct interior scaling enters through the probe insertions: momentum and the singular part of a positive-label mass profile scale as $\tau^{2(p_t-p_x)}$ and $\tau^{2p_t+\sqrt2\ell p_\phi}$. In optical distance their exponents are
\begin{equation}\label{eq:powers}
 q=\frac{2(1-p_x)}{1-p_t},\qquad
 \rho_\ell=\frac{2+\sqrt2\ell p_\phi}{1-p_t}.
\end{equation}
Rescaling by $\omega s$ gives the candidate response powers $\omega^{-q}$ and $\omega^{-\rho_\ell}$. The momentum exponent describes the Kasner-sensitive dispersion, while an ordinary mass response carries the exponent $2/(1-p_t)$~\cite{HartnollZhiboedov}. Distinct spectral structures can therefore recover temporal and spatial metric exponents independently~\cite{XiaoNianLi}. In the $A>1$ Gibbons--Maeda regime, the local $d=3$ metric exponents agree with ours under $A=a$; the asymptotically flat boundary condition and global connection differ. Varying the mass-profile label adds a direction sensitive to the supporting dilaton through $\rho_\ell$.

To connect these local powers to the exterior spectrum, we begin at $a=3$, $Q=\rh=1$, with $\ell_1=\sqrt3$ and $\ell_2=4/\sqrt3$. The predicted exponents are $(q,\rho_{\ell_1},\rho_{\ell_2})=(2/3,2/3,1/3)$. The corresponding terms decay more slowly than regular propagation at order $1/\omega$, so a nonzero projection makes each the leading response.

\subsection{Wave connection and quantization}\label{sec:waveconnection}
A wave connection is the analytic continuation of a solution basis along a specified complex radial path, matching the physical UV and horizon solution lines. We construct it along oscillatory paths, following the monodromy and complex-WKB methods of Refs.~\cite{MotlNeitzke,AnderssonHowls2004,MaassenVanDenBrink2004} and their AdS boundary-matching extensions~\cite{CardosoNatarioSchiappa,NatarioSchiappa2004}. Let $x$ be the continued tortoise coordinate, with $\dd x/\dd r=H^{\gamma/2}/f$, anchored by $x=0$ at the real tip and $x=x_p$ at the UV boundary. Its retarded branch continues the optical logarithms below the outer horizon (appendix~\ref{app:connections}). The tip-to-UV arm and its companion to the outer horizon close with the exterior path into one clockwise turn about the physical horizon. This preserves the ingoing Frobenius line up to a nonzero factor and carries the same source-free condition as the exterior problem.

The connection propagates the two coefficients of $\Psi$ in an order-zero Bessel basis. Set $Z=\omega x$, with its square root and logarithm continued along the chosen optical branch. The normalizable UV solution selects a source-free column; the $+\pi$ optical half-turn carries it to the companion arm, where its outgoing horizon coefficient must vanish. This leaves the retarded ingoing line. Appendix~\ref{app:connections} gives the unit-Wronskian basis, endpoint data and ordered insertion.

Use the normalized source $\mathcal D_{\rm sp}=N_{\rm sp}\mathcal E$, where $\mathcal E$ is the exterior UV source coefficient and $N_{\rm sp}$ is analytic and nonzero on the selected frequency neighbourhoods (appendix~\ref{app:source}). With $\alpha=\omega x_p-\pi/4$, the leading endpoint contraction is
\begin{equation}\label{eq:spacelikeleadingsource}
 \mathcal D_{0,\rm sp}(\alpha)
 =\ii(\cos\alpha+3\ii\sin\alpha),\qquad
 \mathcal D_{0,\rm sp}(\alpha_0)=0.
\end{equation}
Its zero condition yields
\begin{equation}\label{eq:selectedlaw}
 \omega_n^{(0)}
 =\Om_p\left(n+\frac14+\frac{\ii\log2}{2\pi}\right),
 \qquad \Om_p=\frac{\pi}{x_p}.
\end{equation}
The optical action $x_p$ determines global spacing, in the same sense that an analytically continued travel time sets cavity spacing~\cite{GrozdanovMovrinValach}. First-order perturbation about the Bessel equation~\cite{MusiriSiopsisSchwarzschild2003,MusiriNessSiopsis2006} inserts the local power on both arms in physical order. Contracting this complete insertion with the UV and horizon data gives the nonzero physical projection
\begin{equation}\label{eq:physicalprojectionmain}
 \mathcal P(\rho)=\frac{2\pi}{3}I(\rho)
                    (e^{\ii\pi\rho}-1)\ne0,
 \qquad 0<\rho<1,\quad I(\rho)>0.
\end{equation}
The nonzero contraction transmits the local insertion into the exterior pole condition. It links the near-singularity power to a QNM response through the UV and horizon boundary conditions~\cite{HartnollZhiboedov,XiaoNianLi}.

The geometric correction at $a=3$ has exponent $2/3$. UV mass profiles contribute a Coulomb $\log\omega/\omega$ term; regular propagation and horizon matching contribute at order $1/\omega$. For a probe direction $\xi$ with exponent $\rho$, let $G_\xi(\alpha)$ be the coefficient of the leading $\omega^{-\rho}$ source variation. The source and its derivative obey
\begin{equation}\label{eq:spacelikecompletejets}
\begin{aligned}
 \mathcal D_{\rm sp}(\omega,0)
   &=\mathcal D_{0,\rm sp}(\alpha)+O(|\omega|^{-2/3}),\\
 \partial_\xi\mathcal D_{\rm sp}(\omega,0)
   &=\omega^{-\rho}G_\xi(\alpha)
      +O(\log|\omega|/|\omega|).
\end{aligned}
\end{equation}
The logarithm is absent in the momentum remainder. At an actual simple root, the normalization $N_{\rm sp}$ cancels and $-\partial_\xi\mathcal D_{\rm sp}/\partial_\omega\mathcal D_{\rm sp}=-\mathcal E_\xi/\mathcal E_\omega$, as in section~\ref{sec:probe}. The remainder estimates in appendix~\ref{app:source} control the displaced pole and its frequency derivative, giving
\begin{equation}\label{eq:physicalbridge}
 R_\xi(n)=-\frac{G_\xi(\alpha_0)}{x_p\,\partial_\alpha\mathcal D_{0,\rm sp}(\alpha_0)}
                 \omega_n^{-\rho}+o(|\omega_n|^{-\rho}),
 \qquad 0<\rho<1,
\end{equation}
Here $\omega_n$ is the actual pole, $\partial_\alpha\mathcal D_{0,\rm sp}(\alpha_0)\ne0$, and $G_\xi(\alpha_0)$ contains the nonzero local insertion and physical projection. Converting the phase derivative with respect to $\alpha$ into a frequency derivative supplies the factor $x_p$ in the denominator. The three responses therefore decay as $n^{-2/3}$, $n^{-2/3}$ and $n^{-1/3}$. The two equal powers carry different information: momentum probes metric anisotropy, while the mass-label dependence also probes the dilaton.

\subsection{Inverse map and its physical interpretation}
The inverse map expresses the separation of spectral information channels. Mass-label variation disentangles the dilaton weight from temporal metric scaling, while momentum probes metric anisotropy. The mass-label intercept is
\begin{equation}\label{eq:responseintercept}
 \rho_0=\frac{\ell_2\rho_{\ell_1}-\ell_1\rho_{\ell_2}}
              {\ell_2-\ell_1},
\end{equation}
and the metric and dilaton exponents follow from
\begin{equation}\label{eq:inverse}
 p_t=1-\frac2{\rho_0},\qquad
 p_x=1-\frac q{\rho_0},\qquad
 p_\phi=\frac{\sqrt2}{\rho_0}
                \frac{\rho_{\ell_2}-\rho_{\ell_1}}{\ell_2-\ell_1}.
\end{equation}
The intercept $\rho_0=2/(1-p_t)$ removes the label-dependent dilaton weight and isolates the temporal metric exponent $p_t$. It is extrapolated from two nonzero labels, since $W_0=0$ supplies no zero-label response. Once this metric factor is fixed, the slope $\partial_\ell\rho_\ell=\sqrt2p_\phi/(1-p_t)$ measures the dilaton's contribution to the local wave potential and determines its signed exponent in the canonical normalization and chosen coupling orientation. The momentum exponent $q$ then resolves the transverse metric scaling through $p_x=1-q/\rho_0$. Ordinary probe-mass responses already expose the temporal metric exponent~\cite{HartnollZhiboedov,XiaoNianLi}, while also varying the UV mass. The label slope in~\eqref{eq:inverse} supplies the additional matter-sensitive direction at fixed UV dimension.
The Jacobian is $4\sqrt2(\ell_2-\ell_1)/(1-p_t)^4\ne0$, so these response channels distinguish the three exponents without imposing a Kasner constraint in the inversion. The exact EMD triplets themselves lie on the constrained one-parameter curve~\eqref{eq:exactp}. For the example above, $(p_t,p_x,p_\phi)=(-1/5,3/5,-\sqrt6/5)$.

In quantum BTZ, the angular response probes a metric exponent $p_\varphi$, or the combination $2(1-p_\varphi)/(1-p_t)$ when the temporal and spatial factors are independent~\cite{CovielloEtAl,CovielloEtAlLong}. Here $p_\phi$ describes the dilaton supporting the geometry.

\subsection{Propagation terms and numerical reconstruction}\label{sec:spacelikepropagation}
For the $d=3$ spacelike family the local exponents are
\begin{equation}\label{eq:d3powers}
 q=\frac{1+a}{2a}=\frac{1+p_t}{1-p_t},\qquad
 \rho_\ell=1+q-\frac\ell{\sqrt a}.
\end{equation}
The momentum exponent is therefore tied directly to the change in the interior $t$ direction described in section~\ref{sec:geometry}. For a connection through the real spacelike endpoint, $1/3<a<1$ gives $q>1$: the local momentum term decays faster than the regular $1/n$ propagation term. At $a=1$ the two orders coincide, producing the logarithmic resonance
\nopagebreak[4]
\begin{equation}\label{eq:delta1log}
 R_k(n)=\frac1{2\pi}\frac{\log n}{n}+O(n^{-1}).
\end{equation}
For $a>1$, $1/2<q<1$, so the local fractional term decays more slowly than regular propagation. Its nonzero projection makes it the leading momentum response on these spacelike connections. The exponent tends to $1/2$ as the geometry approaches the large-$a$ Kasner limit. Thus $a=1$ marks a change in the ordering of spectral terms within the spacelike regime, whereas $a=1/3$ is the causal threshold of the interior. The role of the real endpoint near this threshold is the subject of section~\ref{sec:critical}.

The subleading regime is represented by $a=1/2$, where $q=3/2$. In $z_n=\omega_n/\Omega_p$,
\begin{equation}\label{eq:halfresponsesmain}
 R_k(n)=\frac3{2\pi z_n}+\mathcal A_kz_n^{-3/2}
                              +O(|z_n|^{-2}\log|z_n|).
\end{equation}
The $1/z_n$ coefficient is the integrated regular momentum contribution; the nonzero $3/2$ term comes from the real endpoint (appendix~\ref{app:connections}). Mass profiles likewise have UV Coulomb logarithms at fixed UV dimension. Separating these propagation terms exposes the local dependence in~\eqref{eq:d3powers}: momentum fixes a metric combination, while the variation with $\ell$ measures the supporting dilaton.

The five backgrounds in table~\ref{tab:spacelikegrades} span this change of response hierarchy. The $a=1/2$ example resolves a local term beneath regular propagation, $a=1$ probes the logarithmic resonance, and $a=3,5,7$ follow the leading fractional regime towards its large-$a$ limit. Across them, the reconstructed $p_t$ passes through zero, $p_x$ increases, and the magnitude of $p_\phi$ decreases, in agreement with the geometric evolution in~\eqref{eq:exactp}.

\begin{table}[!htb]
\centering\small
\setlength{\tabcolsep}{5pt}
\sisetup{round-mode=none}
\begin{tabular}{@{}c *{3}{S[table-format=1.4]}
 S[table-format=-1.4] S[table-format=1.4] S[table-format=-1.4] c@{}}
\toprule
 & \multicolumn{3}{c}{QNM response exponents}
 & \multicolumn{3}{c}{Reconstructed indices} & \\
\cmidrule(lr){2-4}\cmidrule(lr){5-7}
$a$ & {$q$} & {$\rho_{\ell_1}$} & {$\rho_{\ell_2}$}
 & {$p_t$} & {$p_x$} & {$p_\phi$} & {$\Delta_{\max}$}\\
\midrule
$1/2$ & 1.500 & 0.750 & 0.250 & 0.200 & 0.400 & -0.800 & $6.8\times10^{-5}$\\
$1$ & 1.0001 & 0.7500 & 0.2500 & 0.0000 & 0.5000 & -0.7071 & $4.8\times10^{-5}$\\
$3$ & 0.6667 & 0.6667 & 0.3333 & -0.2000 & 0.6000 & -0.4899 & $8.1\times10^{-6}$\\
$5$ & 0.6000 & 0.6000 & 0.2000 & -0.2500 & 0.6250 & -0.3953 & $4.3\times10^{-6}$\\
$7$ & 0.5714 & 0.5714 & 0.1428 & -0.2727 & 0.6364 & -0.3402 & $6.0\times10^{-6}$\\
\bottomrule
\end{tabular}
\caption{Spacelike reconstruction of $(p_t,p_x,p_\phi)$ from three freely extracted QNM response exponents at $Q=\rh=1$, using~\eqref{eq:inverse}. Background-dependent probe labels are listed in table~\ref{tab:probelabels}. The selected-fit discrepancy $\Delta_{\max}=\max_{i=t,x,\phi}|p_i^{\rm QNM}-p_i^{\rm geom}|$ is calculated from the unrounded reconstruction and the geometric indices~\eqref{eq:exactp}; window and remainder-model sensitivity is discussed separately in appendix~\ref{app:numerics}.}\label{tab:spacelikegrades}
\end{table}

\begin{figure}[!htbp]
\centering\includegraphics[width=\linewidth,trim=0 17.0bp 0 0,clip]{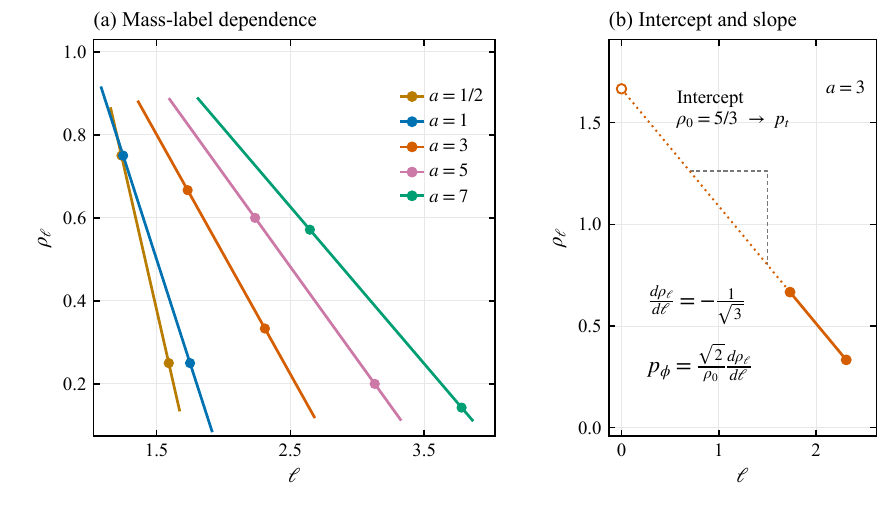}
\caption{Mass-label responses at $Q=\rh=1$. Filled circles are the freely extracted mass-label inputs; solid lines give the geometric predictions. (a) Label dependence for the five backgrounds. (b) Intercept and slope at $a=3$: the intercept fixes $p_t$, and together with the slope determines the signed $p_\phi$ through~\eqref{eq:inverse}. The dotted extension and open circle at $\ell=0$ show the extrapolated intercept.}\label{fig:probe-exponents}
\end{figure}

Figure~\ref{fig:probe-exponents}(a) shows how the mass-label slope $-1/\sqrt a$ becomes less steep as $a$ increases. The two mass-label inputs are shown for each of the five backgrounds; additional resolved labels support the same relation beyond those inputs. Panel~(b) makes the two information channels visible at $a=3$: the extrapolated intercept $\rho_0=5/3$ gives $p_t=-1/5$, and combining it with the slope $-1/\sqrt3$ gives the signed dilaton exponent $p_\phi=-\sqrt6/5$ through~\eqref{eq:inverse}.

\begin{figure}[!htbp]
\centering\includegraphics[width=\linewidth,trim=0 11.7bp 0 0,clip]{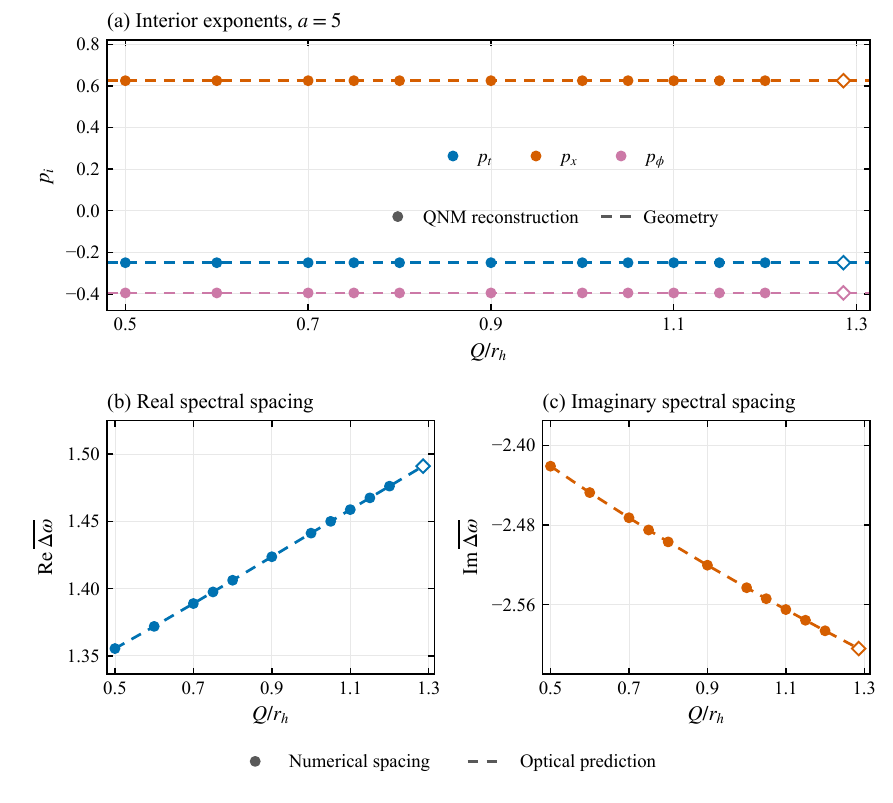}
\caption{Local scaling and global spectral spacing at $a=5$, $\rh=1$. (a) Reconstructed interior indices compared with the geometric constants~\eqref{eq:exactp}. (b,c) Real and imaginary mean spacings compared with the optical prediction $\Omega_p=\pi/x_p$. Here $\overline{\Delta\omega}$ averages $\omega_{n+1}-\omega_n$ over $n=280,\ldots,299$. Open diamonds mark $Q/\rh=9/7$, where the ratio $T/\mu$ reaches its minimum.}\label{fig:charge-kasner}
\end{figure}

\FloatBarrier
Figure~\ref{fig:charge-kasner} separates local scaling from global propagation. At fixed $a=5$, the reconstructed triplet in panel~(a) stays unchanged as charge varies, whereas the real and imaginary spacings in panels~(b,c) change with the state. The exponents~\eqref{eq:exactp} depend only on $a$, while $\Omega_p=\pi/x_p$ in~\eqref{eq:selectedlaw} samples the whole geometry. The twelve states follow both relations and include $Q/\rh=9/7$, where $T/\mu$ is minimal along the $a=5$ family. Here $T$ is the Hawking temperature and $\mu$ the boundary chemical potential; their background dependence is given in Eqs.~(2.12)--(2.14) of Ref.~\cite{AreanEtAl}.

The Gibbons--Maeda analysis~\cite{XiaoNianLi} likewise separates state-dependent global spectral data from fixed local metric exponents. Here the local data also include the dilaton slope, while the charge dependence of spacing distinguishes states with the same interior scaling. The lower-charge state $Q/\rh=2/5$ is excluded because its finite-window reconstruction remains sensitive to the momentum remainder template (appendix~\ref{app:numerics}). In this reconstruction the real spacelike endpoint supplies the local response powers. The critical background provides a different connection, allowing us to examine how another endpoint changes their spectral imprint.

\FloatBarrier
\section{Critical and near-critical interiors}\label{sec:critical}
\subsection{Local scaling at the real and complex endpoints}
At criticality, the real and complex endpoints have opposite dilaton limits and distinct local response hierarchies. Their comparison makes the role of the wave connection explicit: local scaling supplies possible spectral powers, but the connection determines which endpoint contributes. For $a=1/3$, $Q=\rh=1$, define $H=1+1/r$. Combining the two blackening terms gives
\begin{equation}\label{eq:criticalrealgeometry}
f=\frac{(r+1)^3-8}{r}=-\frac7r+3+3r+r^2,\qquad
 \phi=-\frac{\sqrt3}{2}\log H.
\end{equation}
The real singularity remains spacelike, with $\tau\propto r^{3/4}$ and $(p_t,p_x,p_\phi)=(1/3,1/3,-\sqrt{2/3})$. Ordinary momentum has exponent two and a constant leading potential, so it does not supply a distinct fractional metric response. For mass labels $\ell_1=5/(2\sqrt3)$, $\ell_2=9/(4\sqrt3)$ and gradient label $\ell_g=\sqrt3/2$, the candidate real-tip exponents are
\begin{equation}\label{eq:criticalgradientgrade}
 \rho_\ell=3-\sqrt3\ell,\qquad
 \rho_g=2-\sqrt3\ell_g
 =\frac{2(1-p_x)+\sqrt2\ell_g p_\phi}{1-p_t},
\end{equation}
namely $(\rho_{\ell_1},\rho_{\ell_2},\rho_g)=(1/2,3/4,1/2)$. The two mass directions and the gradient direction have a locally invertible metric--dilaton power map,
\begin{equation}\label{eq:criticallocalrank}
 \det\frac{\partial(\rho_{\ell_1},\rho_{\ell_2},\rho_g)}
               {\partial(p_t,p_x,p_\phi)}
 =\frac{4\sqrt2(\ell_2-\ell_1)}{(1-p_t)^4},
\end{equation}
so the mass and gradient channels retain local dilaton sensitivity even where ordinary momentum loses its fractional response~\cite{HartnollZhiboedov}.

The analytic wave equation also has a distinguished complex endpoint: $H=0$, or $r=-1$, separate from the real singularity $H=\infty$. Use one continued logarithm with $-\pi<\arg H<\pi$; here the profile powers are $b_\ell=\sqrt3\ell/2$ and $b_{\ell_g}=\sqrt3\ell_g/2$. The optical equation and its endpoint expansion are
\begin{equation}\label{eq:criticalexactoptical}
\begin{gathered}
 \mathcal G=-(H-2)(7H^2-10H+4),\\[3pt]
 \frac{\dd X}{\dd H}=-\frac{H^{3/2}}{\mathcal G},\qquad
 X=-\frac{H^{5/2}}{20}[1+O(H)].
\end{gathered}
\end{equation}
The full wave insertions are
\begin{equation}\label{eq:criticalcomplexinsertions}
 U_\ell=-\mathsf D+\mathsf D H^{b_\ell},\qquad
 U_g=\frac{\mathsf D}{\mathsf C}H^{b_{\ell_g}}.
\end{equation}
At the real tip, $H\to\infty$ and $\phi\to-\infty$, so the positive-label exponential grows and distinguishes the mass labels. At the complex endpoint, $H\to0$ and $\Re\phi\to+\infty$: both mass profiles approach $W_\ell\to-1$, while the gradient profile approaches zero. The resulting endpoint expansions are
\begin{equation}\label{eq:criticalendpointinsertions}
\begin{aligned}
 U_\ell&=-8H^{-3/2}[1+O(H)]
           +8H^{b_\ell-3/2}[1+O(H)],\\
 U_g&=8H^{b_{\ell_g}-3}[1+O(H)].
\end{aligned}
\end{equation}
The common mass term comes from the $-1$ in the UV-preserving profile. At the UV, $H\to1$ cancels the leading mass terms; at the complex endpoint, the exponential piece vanishes relative to $-\mathsf D$. The same subtraction that fixes the UV dimension therefore produces a common singular mass insertion at this endpoint. For the gradient, the stronger metric weight $\mathsf D/\mathsf C$ makes the insertion singular even though its dilaton-dependent profile vanishes. Using the optical expansion, the local response exponents associated with insertions $X^{\rho-2}$ are
\begin{equation}\label{eq:criticalprofilegrades}
\rho_c=\frac75,\quad
 \rho_{c,g}=(\rho_c-1)(2+b_{\ell_g})=\frac{11}{10},\quad
 \rho_c+(\rho_c-1)b_\ell=\frac{7+2b_\ell}{5}.
\end{equation}
Thus the first mass layer has exponent $7/5$ for both labels, their later layers have exponents $19/10$ and $37/20$, and the gradient has exponent $11/10$. The opposite dilaton limits at the two endpoints produce different local response hierarchies. To identify the hierarchy seen by the exterior poles, we now follow the wave connection from the UV boundary to the horizon.

\subsection{Complex-endpoint connection and quantization}
\begin{figure}[!htbp]
\centering\includegraphics[width=\linewidth,trim=0 25.8bp 0 0,clip]{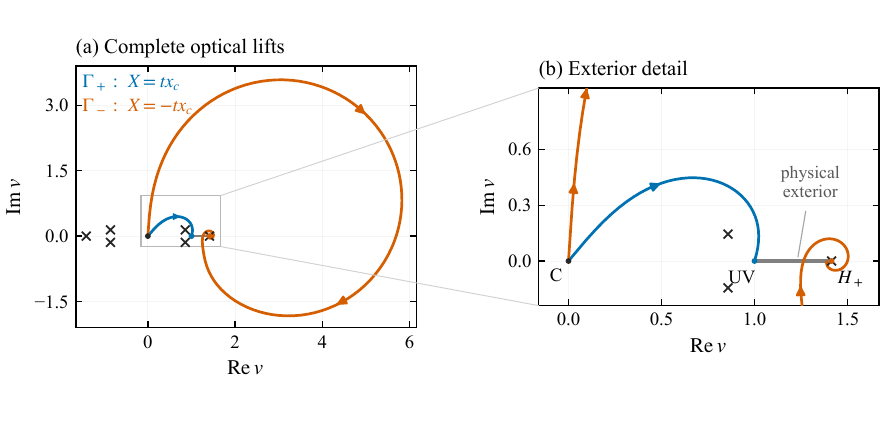}
\caption{Optical connections at criticality, $a=1/3$, $Q=\rh=1$, in $v=\sqrt H$. (a) Complete paths; (b) the boxed region enlarged. The complex endpoint $C$, UV boundary and outer horizon $H_+$ lie at $v=0,1,\sqrt2$, respectively; the grey segment $1\le v\le\sqrt2$ is the physical exterior. The arms $\Gamma_\pm$ follow $X=\pm tx_c$ to the UV and outer horizon, respectively, where $X$ is the optical coordinate and $t$ the path parameter. Arrows indicate increasing $t$; crosses mark optical poles.}\label{fig:criticaloptical}
\end{figure}

We apply the AdS Bessel matching of section~\ref{sec:waveconnection}~\cite{CardosoNatarioSchiappa,NatarioSchiappa2004} to the competing complex endpoint. The physical exterior lies between $H=1$ and $H=2$. The UV action measured from the complex endpoint is fixed by the upper optical pole residue $\mathcal R_c$:
\begin{equation}\label{eq:criticalactiondefinition}
 x_c=-2\pi\ii\mathcal R_c
     -\int_0^1\frac{H^{3/2}}{\mathcal G(H)}\,\dd H .
\end{equation}
The exact residues and boundary signs in appendix~\ref{app:criticalpath} give
\begin{equation}\label{eq:criticalactionsigns}
 \Re x_c>0,\qquad 0<\Im x_c<\pi\mathcal R_h .
\end{equation}
On the principal-sheet germs, the positive optical arm $X=tx_c$ reaches the UV at $t=1$, while its $+\pi$ companion $X=-tx_c$ reaches the outer-horizon logarithmic cap. The first-contact and Jordan-domain argument in that appendix establishes these endpoint connections analytically. Numerical integration then constructs the inverse branches shown in figure~\ref{fig:criticaloptical} relative to the real exterior. The half-turn occurs in the $X$ plane: since $X\sim-v^5/20$ near $v=0$, it corresponds locally to a $\pi/5$ separation of the lifted rays in the $v$ plane.

The connection continues the two physical endpoint solutions around
$H=0$: the positive arm carries the UV-normalizable solution and the
negative arm reaches the outer-horizon ingoing solution.
Appendix~\ref{app:criticalpath} constructs a pole-free domain between this
path and the real exterior. Once the endpoint caps are fixed, the only
possible residual loop surrounds the outer horizon. Its monodromy
preserves the ingoing Frobenius line and multiplies it by a nonzero
analytic factor. Consequently the exact source obtained by continuation
satisfies
\begin{equation}\label{eq:criticalretardedequivalence}
 \mathcal E^{\rm cont}=M_H\mathcal E^{\rm ext},\qquad
 \mathcal D=\mathcal N\mathcal E^{\rm ext},\qquad
 \mathcal N\ne0,
\end{equation}
where $M_H$ and $\mathcal N$ are analytic on the nonresonant frequency
domains used below. The second factor includes the normalization of the
coefficient-space source. The exact $\mathcal D$ therefore has the exterior
QNM zeros and the same simple-pole response,
$-\mathcal D_\xi/\mathcal D_\omega
=-\mathcal E^{\rm ext}_\xi/\mathcal E^{\rm ext}_\omega$:
derivatives of $\mathcal N$ multiply the vanishing source. The Bessel
contraction below approximates this exact source, with its error and first
variation controlled in appendix~\ref{app:source}. No additional boundary
condition is imposed at the real singularity, the complex endpoint, or an
inner horizon.

The geometric potential at $H=0$ has Bessel index $\nu_c=1/5$. Retaining both $J_{\nu_c}$ and $Y_{\nu_c}$ columns gives the half-turn $T_+(\nu_c)$ in~\eqref{eq:complexhalfturn}.
The UV regular vector is $(\cos A_\omega,\sin A_\omega)^T$, where $A_\omega=\omega x_c-7\pi/20$; the row $(1,\ii)$ extracts the forbidden outgoing component on the negative arm. Their contraction gives the leading normalized source
\begin{equation}\label{eq:criticalphysicalsource}
 \mathcal D_0=-2\ii\cos(\pi/5)e^{2\ii\omega x_c}-1,
\end{equation}
and hence
\begin{equation}\label{eq:complexleadingroots}
\begin{aligned}
 \omega_n^{(0)}&=\Omega_c(n+\beta_c),\\
 \Omega_c&=\frac\pi{x_c},\qquad
 \beta_c=\frac14+\frac{\ii}{2\pi}\log[2\cos(\pi/5)].
 \end{aligned}
\end{equation}
The complex-endpoint spacing differs substantially from that based on the alternative real-tip action,
\begin{equation}\label{eq:criticalgap}
 \Omega_p=\pi/x_p\simeq1.0113-3.9861\ii,\qquad
 \Omega_c=\pi/x_c\simeq1.5889-4.9037\ii .
\end{equation}
The numerical tower tracked through $n=640$ has spacing consistent with $\Omega_c$. The complete source estimates in appendix~\ref{app:source} construct the corresponding asymptotic simple-root family.

\FloatBarrier
\Needspace{8\baselineskip}
\subsection{Response hierarchy at criticality}
At criticality, regular propagation gives the leading inverse-frequency terms, including the mass probe's UV Coulomb logarithm. Complex-endpoint scattering supplies the smaller fractional-power terms and subsequent corrections. For $1<\rho<2$ and $\rho>2\nu$ (here $\nu=\nu_c=1/5$), a unit local insertion $X^{\rho-2}$ acquires a factor $\omega^{-\rho}$ when $Z=\omega X$. The complete two-column Bessel kernel $\mathcal R_\rho$ has large-$Z$ mean $Z^{\rho-2}J/2$, where $J=\left(\begin{smallmatrix}0&1\\-1&0\end{smallmatrix}\right)$. At a fixed matching point $X_0$ in the local overlap, subtracting this mean defines the finite part $H_{\rm FP}$ of~\eqref{eq:timelikeFPdefinition} and gives
\begin{equation}\label{eq:localinsertionfrequencysplit}
\omega^{-\rho}\int_0^{\omega X_0}\mathcal R_\rho(Z)\,\dd Z
 =\frac{X_0^{\rho-1}}{2(\rho-1)\omega}J
 +\omega^{-\rho}H_{\rm FP}+o(|\omega|^{-\rho}).
\end{equation}
All powers follow the continued optical branch, with the companion arm obtained by the $+\pi$ half-turn. The integrated mean produces a regular $1/\omega$ term, while the finite part preserves the fractional power $\omega^{-\rho}$. This separation parallels the regular and non-analytic response terms isolated in Ref.~\cite{HartnollZhiboedov} and the overlap subtraction in appendix A of Ref.~\cite{XiaoNianLi}.

The remaining propagation contribution is fixed by UV matching. For the mass probe, the Coulomb tail $b_\ell/y$, with $y=x_c-X$, adds $\Log\omega/\omega$; the bounded gradient insertion has no such logarithm. This UV effect differs from the near-singularity background logarithms discussed in section 3.2 of Ref.~\cite{HartnollZhiboedov}. Boundary geodesic correlators likewise distinguish a UV logarithm from a singularity-dependent nonanalytic term~\cite{FrenkelEtAl}; here their separation follows from the wave insertion and UV matching. Restoring the mean on the rest of the arms cancels the artificial matching point $X_0$. The UV-side phase fixes the regular coefficients, whereas the horizon-side common amplitude cancels from the response at a pole.

The pole derivative $R_\xi(n)=-[\mathcal D_\xi/\mathcal D_\omega]_{(\omega_n,0)}$ combines these propagation and endpoint contributions; appendix~\ref{app:source} supplies the source estimates. With $z=\omega/\Omega_c$, the mass regular coefficients are
\begin{equation}\label{eq:criticallabelregular}
 B_i=\frac{b_{\ell_i}}{2\pi},\qquad
 C_i=B_i[\Log(-2\Omega_c)-\psi(b_{\ell_i})]-\frac1{2\pi}.
\end{equation}
The gradient mean gives $-2/(\pi z)$. The fractional layers follow from the two-arm projection $A(\rho,c_{\mathrm{loc}})$ in appendix~\ref{app:coefficients}. The nonzero coefficients $A_m=A(\rho_c,-8)=A(7/5,-8)$ and $A_g=A(\rho_{c,g},8)=A(11/10,8)$ give
\begin{subequations}\label{eq:criticalsurvivingresponse}
\begin{align}
 R_{\ell_i}&=\frac{B_i\Log z+C_i}{z}
               +A_mz^{-\rho_c}+O(|z|^{-9/5}),
 \label{eq:criticalsurvivingresponse:mass}\\
 R_g&=-\frac2{\pi z}+A_gz^{-\rho_{c,g}}+O(|z|^{-3/2}).
 \label{eq:criticalsurvivingresponse:gradient}
\end{align}
\end{subequations}
Both mass channels share $A_m$ because their profiles have the same limiting value at the participating endpoint. Their complete responses retain dilaton-coupling label dependence through $B_1-B_2=1/(16\pi)$ and through the later endpoint exponents. The common $7/5$ layer records the shared profile limit. Relative to the real-endpoint response, the complex connection distributes matter sensitivity among different spectral orders.

The complete mass and gradient responses are therefore $O(\log n/n)$ and $O(n^{-1})$, respectively, also following from the source estimates in appendix~\ref{app:source}. Comparing them with the real-tip candidate powers at $z_n=\omega_n/\Omega_c$ gives
\begin{equation}\label{eq:criticalrepresentativeabsence}
 z_n^{1/2}R_{\ell_1}\longrightarrow0,\qquad
 z_n^{3/4}R_{\ell_2}\longrightarrow0,\qquad
 z_n^{1/2}R_g\longrightarrow0.
\end{equation}
All three candidate exponents are below one, so nonzero terms with these powers would decay more slowly than the complete responses. Equation~\eqref{eq:criticalrepresentativeabsence} excludes them as leading terms of this spectral family. Local dilaton sensitivity at the real tip thus need not become a corresponding pole response: the wave connection selects which endpoint information enters the exterior spectrum. Regular-black-hole examples likewise show that complex singularities and the trajectory of asymptotic solutions can control the spectrum even when the real centre is regular~\cite{LanWang2023}.

This endpoint selection clarifies the connection requirement in singularity-sensitive response formulae. The formulae in Ref.~\cite{CovielloEtAlLong} require the real singularity to participate in the dominant connection with nonzero leading projection. Here a different endpoint controls the response despite the nondegenerate local mass--gradient map at the real tip. Complex sectors also appear in Ref.~\cite{XiaoNianLi}; in that Gibbons--Maeda analysis the remote contribution is exponential in frequency, whereas the present endpoint sets an algebraic response hierarchy and changes the information available for inversion.

\begin{figure}[!htbp]
\centering\includegraphics[width=\linewidth,trim=0 5.2bp 0 0,clip]{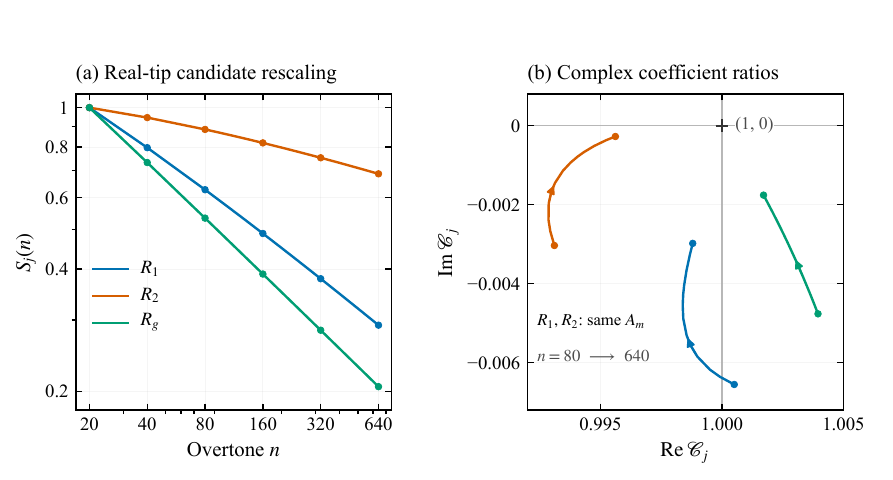}
\caption{Critical responses at $a=1/3$, $Q=\rh=1$, plotted as in~\eqref{eq:criticalplotquantities}. (a) Real-tip-rescaled magnitudes $S_j$, each normalized by its own $n=20$ value. (b) Complex coefficient ratios $\mathcal C_j$ after regular subtraction, with theoretical target $(1,0)$. Lines join all fifteen samples in (a) and the thirteen with $n\ge80$ in (b); filled circles mark six selected overtones in (a) and the endpoints $n=80,640$ in (b). Arrows indicate increasing $n$. Panel (b) uses different horizontal and vertical scales.}\label{fig:criticalresponses}
\end{figure}

Figure~\ref{fig:criticalresponses} compares the candidate real-tip scaling with the complex-endpoint coefficients. At the numerical frequencies $z_n=\omega_n/\Omega_c$, with $R_i=R_{\ell_i}$, the plotted quantities are
\begin{subequations}\label{eq:criticalplotquantities}
\begin{align}
 S_j(n)&=\frac{|z_n^{\rho_j}R_j(z_n)|}
 {|z_{20}^{\rho_j}R_j(z_{20})|},\label{eq:criticalplotS}\\
 \mathcal C_j(z_n)&=\frac{z_n^{\rho_{c,j}}
 [R_j(z_n)-R_{j,\mathrm{reg}}(z_n)]}{A_j}.
 \label{eq:criticalplotC}
\end{align}
\end{subequations}
Here $(\rho_1,\rho_2,\rho_g)=(1/2,3/4,1/2)$, while
$\rho_{c,1}=\rho_{c,2}=\rho_c=7/5$ and $\rho_{c,g}=11/10$.
The regular terms and fixed coefficients are those of~\eqref{eq:criticalsurvivingresponse}:
$A_1=A_2=A_m$ and the gradient coefficient is $A_g$.
All powers and the regular subtraction use the same continued logarithm at the actual complex frequencies.

The decrease of $S_j(n)$ over the displayed range is consistent with the absence of the slower real-tip powers in~\eqref{eq:criticalrepresentativeabsence}. Removing the larger regular terms in panel~(b) reveals the approach to the shared fractional mass coefficient, while the gradient follows its distinct endpoint prediction. The approach is not strictly monotonic: $|\mathcal C_2-1|$ increases slightly from $n=80$ to $90$, then decreases at the later sampled overtones. The exterior numerical responses resolve these leading coefficients. The smaller mass difference probes later label-dependent terms that remain quantitatively unresolved (appendix~\ref{app:numerics}).

\FloatBarrier
\subsection{Continuation across the causal threshold}\label{sec:localcontinuation}
The critical mechanism is already present without an inner horizon. To follow it across the causal threshold, keep each physical mass or gradient label fixed; its profile power $b_\ell=2\ell\sqrt a/(1+a)$ still varies with the background. The connection, optical action $x_c$ and endpoint phase offset $\beta_c$ continue in a two-sided neighbourhood of $a=1/3$. The size of this neighbourhood depends on the labels; appendix~\ref{app:localcontinuation} gives their allowed ranges and controls the continued source.

For this continued family the leading spectrum and response orders are
\begin{equation}\label{eq:localcontinuationspectrum}
\begin{aligned}
 \omega_n&=\frac{\pi}{x_c}\bigl[n+\beta_c+O(n^{-\gamma_*})\bigr],
               \qquad \gamma_*>0,
 \\
 R_\ell(n)&=O(\log n/n),\qquad R_g(n;\ell_g)=O(n^{-1}).
\end{aligned}
\end{equation}
Here $x_c$ and $\beta_c$ follow the continued background, while $\gamma_*>0$ is the locally uniform remainder exponent controlled in appendix~\ref{app:localcontinuation}; $z_n=\omega_nx_c/\pi$. The real-tip reference exponents remain between zero and one, so
\begin{equation}\label{eq:localcontinuationabsence}
 z_n^{\rho_m}R_\ell\longrightarrow0,\qquad
 z_n^{\rho_g}R_g\longrightarrow0.
\end{equation}
The family retains its regular response orders across the threshold, with the slower real-tip powers absent. On the timelike side these reference exponents describe proper-radial scaling.

As $a$ approaches $1/3$ from the timelike side, the newly formed inner horizon moves to $H=\infty$, beyond the finite domain used to continue the complex-endpoint arms. The real singularity can therefore change its causal character while the participating complex endpoint, wave connection and leading response mechanism remain continuous.

This continuity in the background parameter differs from the overtone-resolved real Kasner transitions of Ref.~\cite{HartnollZhiboedov} and from the charged quantum-BTZ crossover, whose oscillatory approach has a conjectured inner-horizon origin~\cite{CovielloEtAlLong}. The endpoint mechanism can therefore persist while the real interior changes causal character. To examine timelike interiors away from this neighbourhood, we next construct their wave connections and resolve the associated response layers.

\section{Timelike interiors}\label{sec:timelike}
In a timelike interior, the local radial scaling and the endpoint reached by the wave connection play distinct roles in the spectrum. We examine this relation through analytic response coefficients at five known backgrounds. Their numerical comparison is a forward test; the model-dependent inverse map requires an independently measured endpoint exponent.

\subsection{Radial scaling and wave connection}
For $0<a<1/3$, the region beyond the inner horizon ends at a timelike singularity. Proper distance $\sigma$ measures a spatial approach to that singularity: the metric scales with exponent $a$ and the supporting dilaton with slope $2\sqrt a$, as in~\eqref{eq:timelikepowers}. These are radial scalings, rather than temporal contraction or stretching along a Kasner evolution. For the five backgrounds at $Q=\rh=1$, choose
\begin{equation}\label{eq:timelikeprofile}
 \ell\delta=\frac12+a,\qquad
 b_\ell=\frac{1+2a}{1+a},\qquad W_\ell=H^{b_\ell}-1 .
\end{equation}
The response $R_\ell$ is evaluated at known $a$, holding the background and its chosen label fixed in each strength derivative. These labels vary between backgrounds, in contrast to the fixed-label continuation in section~\ref{sec:critical}. At the real radial tip, $H\to\infty$ and the profile grows, giving the reference exponent
\begin{equation}\label{eq:subcriticalrealgrade}
 \rho_p=\frac{1-2a}{1-a}<1 .
\end{equation}
Along this label curve, the timelike mass reference~\eqref{eq:localrealtipgrades} stays in the range $1/2<\rho_p<1$ throughout $0<a<1/3$. A transmitted real-tip contribution would decay more slowly than regular $\log n/n$ propagation. At the participating complex endpoint, however, $b_\ell>0$ gives $W_\ell\to-1$. This shared limit produces a common leading fractional mass layer for positive labels at a fixed background. The labels remain distinguishable through the regular Coulomb and finite-phase terms and the later $H^{b_\ell}$ endpoint layers.

At $a=1/9,1/7,1/5,1/4,2/7$, we construct the complex-endpoint connections separately. Each comprises two optical arms, their $+\pi$ half-turn, a fixed UV reference sheet and an outer-horizon cap. The finite framed comparison reduces to two clockwise outer-horizon turns, preserving the ingoing line and hence the physical boundary condition. Appendix~\ref{app:connections} gives the geometric construction.

The complex-endpoint response exponent and Bessel index are
\begin{equation}\label{eq:subcriticalgeneralgrades}
 \rho_c=\frac{2(2+a)}{3+a}>1,\qquad
 \nu_c=\frac{\rho_c-1}{2}.
\end{equation}
Contracting the UV regular column with $T_+(\nu_c)$ and the outgoing row gives
\begin{subequations}\label{eq:timelikequantization}
\begin{align}
 \mathcal D_0&=-2\ii\cos(\pi\nu_c)e^{2\ii\omega x_c}-1,
 \label{eq:timelikequantization:source}\\
 \omega_n^{(0)}&=\frac\pi{x_c}
 \left(n+\frac14+\frac{\ii}{2\pi}\log[2\cos(\pi\nu_c)]\right).
 \label{eq:timelikequantization:frequencies}
\end{align}
\end{subequations}
The action $x_c$ includes the residue-log branches of each connection. Its nonzero value, the ingoing cap margin and the geometric exponent $2(\rho_c-1)$ enter the simple-root argument of appendix~\ref{app:source}. The UV Coulomb tail and complex-endpoint exponents greater than one then set the response order. At each background, writing $\Omega_c=\pi/x_c$ and $z_n=\omega_n/\Om_c$, this pole family satisfies
\begin{equation}\label{eq:fullintervalresponsebound}
\begin{aligned}
 R_\ell(n)&=O_a(\log n/n),\\
 |z_n^{\rho_p}R_\ell(n)|
 &=O_a\!\left(\frac{\log n}{n^{a/(1-a)}}\right)\longrightarrow0.
\end{aligned}
\end{equation}
The total leading real-tip coefficient therefore vanishes at these five backgrounds, extending the complex-endpoint picture beyond the local critical neighbourhood.

\subsection{Regular and fractional responses}
The full insertion of the profile~\eqref{eq:timelikeprofile} obeys the exact one-form
\begin{equation}\label{eq:subcriticaloneform}
 U_\ell\,\dd X=
 \frac{1-H^{b_\ell}}{(1-H)^2}\dd H .
\end{equation}
Its positive-arm primitive has a logarithmic UV part. Matching its restored mean to the UV Coulomb phase cancels the cutoff and gives $F(\omega)=b_\ell[\Log(-2\omega)-\psi(b_\ell)]-1$. With $z=\omega/\Omega_c$ and $\Omega_c=\pi/x_c$, the mean/finite-part decomposition~\eqref{eq:localinsertionfrequencysplit} therefore fixes
\begin{equation}\label{eq:subcriticalregular}
 B_\ell=\frac{b_\ell}{2\pi},\qquad
 C_\ell=B_\ell[\Log(-2\Om_c)-\psi(b_\ell)]-\frac1{2\pi}.
\end{equation}
The logarithm is fixed by the UV Coulomb tail, while the constant contains the finite arm phase. At the complex endpoint, $X\propto H^{(3+a)/(1+a)}$ and the constant profile limit $W_\ell\to-1$ gives $U_\ell\sim\alpha_cX^{\rho_c-2}$, with
\begin{equation}\label{eq:complexlocalamplitude}
 \alpha_c=2^{2(2-\rho_c)}
           (\rho_c-1)^{2-\rho_c}e^{\ii\pi(\rho_c-1)}.
\end{equation}
Combining regular propagation with this fractional layer gives the complete response
\begin{equation}\label{eq:subcriticalcomplete}
 R_\ell(z)=\frac{B_\ell\Log z+C_\ell}{z}
       +A_cz^{-\rho_c}
       +O_a(|z|^{1-2\rho_c}\log|z|),
\end{equation}
with remainder exponent $2\rho_c-1=(5+3a)/(3+a)$. Its fixed fractional amplitude and phase are
\begin{equation}\label{eq:subcriticalAcclosed}
 A_c=\mathcal F(\rho_c-1)
       \exp\!\left[(\rho_c-1)
          \left(\frac{3\pi\ii}{2}-\Log\Om_c\right)\right],
\end{equation}
where $\mathcal F$ is the Gamma prefactor~\eqref{eq:timelikeGammaprefactor} derived in appendix~\ref{app:coefficients}. The nonzero two-arm projection carries the local fractional insertion into the retarded response.

Since $\rho_c>1$, regular propagation dominates at high overtone number and the subleading fractional layer records the complex endpoint. The regular coefficients retain the dilaton-coupling label through $b_\ell$, even though the spectator's UV dimension is fixed. The response therefore combines propagation with endpoint scattering, rather than following the slower real-tip power~\eqref{eq:subcriticalrealgrade}.

\suppressfloats[t]
\subsection{Spectral responses and a model-dependent inverse map}
Within this analytic EMD model, the complex-endpoint exponent $\rho_c=2(2+a)/(3+a)$ and the real proper-radial geometry depend on the same background parameter. This exponent is monotonic, with
\begin{equation}\label{eq:timelikemodelinverse}
 a=\frac{3\rho_c-4}{2-\rho_c}.
\end{equation}
An independently measured $\rho_c$ would therefore constrain the radial metric power $a$ and dilaton slope $2\sqrt a$ through the model's relation between the complex endpoint and the real geometry. Unlike the spacelike inverse, this relation uses a single response exponent to infer quantities linked by the background solution. The comparison below tests the predicted coefficients at known backgrounds.

\begin{figure}[!htbp]
\centering\includegraphics[width=\linewidth,trim=0 20.4bp 0 0,clip]{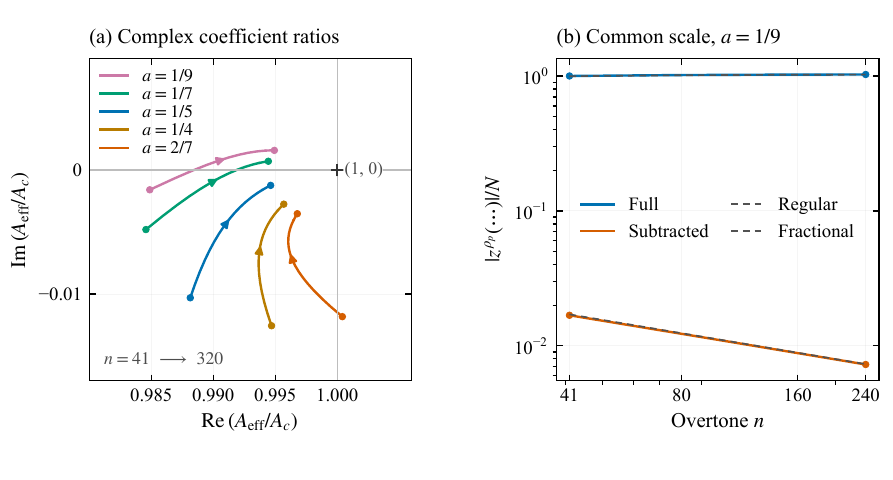}
\caption{Timelike responses at $Q=\rh=1$ for the probes~\eqref{eq:timelikeprofile}. (a) Coefficient ratios $A_{\mathrm{eff}}/A_c$: solid tracks use every integer $n=41,\ldots,240$, and dashed guides connect the additional points at $n=280,320$. Filled circles mark the endpoints $n=41,320$, arrows indicate increasing $n$, and the cross marks the theoretical target $(1,0)$. (b) Magnitudes with the common normalization~\eqref{eq:timelikeplotmagnitudes}, at $a=1/9$ and $n=41,\ldots,240$. Coloured solid curves show the full and subtracted numerical responses; grey dashed curves give the regular and fractional predictions.}\label{fig:subcriticalcomplete}
\end{figure}

Define the subtracted effective coefficient
\begin{equation}\label{eq:effectivecoefficient}
 A_{\rm eff}(z)=z^{\rho_c}
 \left[R_\ell(z)-\frac{B_\ell\Log z+C_\ell}{z}\right].
\end{equation}
The fixed prediction is $A_{\rm eff}/A_c\to1$. Figure~\ref{fig:subcriticalcomplete}(a) compares its amplitude and phase at the five known backgrounds, using responses at their actual complex frequencies. Panel~(b) puts the full, regular, subtracted and fractional contributions on a common scale. At $a=1/9$, set $R=R_\ell$; the four plotted magnitudes are
\begin{equation}\label{eq:timelikeplotmagnitudes}
\begin{gathered}
 \frac{|z^{\rho_p}R|}{N},\qquad
 \frac{|z^{\rho_p}R_{\mathrm{reg}}|}{N},\qquad
 \frac{|z^{\rho_p}(R-R_{\mathrm{reg}})|}{N},\qquad
 \frac{|A_c z^{\rho_p-\rho_c}|}{N},\\[3pt]
 N=|z_{41}^{\rho_p}R(z_{41})|,\qquad
 R_{\mathrm{reg}}=\frac{B_\ell\Log z+C_\ell}{z}.
\end{gathered}
\end{equation}
Here $\rho_p=7/8$ and $\rho_c=19/14$. The subtraction is performed in the complex plane before taking the modulus, with the same logarithmic branch throughout. These magnitudes need not add to the full magnitude.

The common scale in figure~\ref{fig:subcriticalcomplete}(b) shows that regular propagation dominates the full response, while the smaller subtracted contribution follows the fixed fractional prediction. After real-tip rescaling, the regular terms behave as $n^{-1/8}(B_\ell\log n+C_\ell)$ at leading order. Their slow variation over the displayed interval accounts for the nearly flat full response. The isolated complex-endpoint term has the asymptotic power $n^{\rho_p-\rho_c}=n^{-27/56}$; its smaller magnitude and faster decay explain why it is clearer after regular subtraction.

\Needspace{5\baselineskip}
The complex distances $|A_{\rm eff}/A_c-1|$ decrease from $n=240$ to $280$ and $320$ at all five backgrounds, strengthening the numerical trend towards the analytic limit. The smaller fractional coefficient is resolved less accurately than the full response, which is dominated by the regular term. Appendix~\ref{app:numerics} quantifies both comparisons.

The contrast with charged quantum BTZ illustrates the role of the probe and participating endpoint~\cite{CovielloEtAlLong}. There the angular response $n^{-2/3}$ reflects the charge-dominated scaling at the real timelike singularity. Here the response to $W_\ell=H^{b_\ell}-1$ combines a dominant regular term with a fractional layer from $H=0$. The causal character of the real singularity alone therefore does not specify the response hierarchy: the probe weight and its wave connection determine how interior scaling appears in the spectrum.

\FloatBarrier
\section{Discussion}\label{sec:discussion}
High-overtone responses distinguish metric and supporting-matter information through separate channels. For the spacelike real-endpoint connections, the mass-label intercept isolates temporal metric scaling. Combining it with the slope resolves the signed dilaton exponent, while momentum supplies the transverse metric information. The inverse map expresses this separation at fixed UV dimension without using a Kasner constraint. Early power-law analyses related angular metric scaling to the leading spectral condition~\cite{DasShankaranarayanan2005,GhoshShankaranarayananDas2006}, and momentum and mass responses now resolve metric exponents through distinct subleading structures~\cite{HartnollZhiboedov,XiaoNianLi}. The dilaton-coupling label adds a direction that directly probes the matter supporting the geometry.

The probe label converts the dilaton's logarithmic profile into a changing response power at fixed UV dimension. Other observables recover complementary combinations of local data. For finite-polynomial blackening factors shared by temporal and radial metric components in even boundary dimensions, zero-spatial-separation OPE fits relate the complex-time bouncing power to the temporal Kasner exponent when the leading inverse-radius power is higher than quadratic~\cite{CeplakValach2026}. In the EMD backgrounds~\cite{AreanEtAl}, the constant-mean-curvature rate in Eq.~(4.42) is proportional to $\sqrt{1-p_\phi^2}$. The thermal $a$-function power in Eq.~(4.54) determines $p_t$ and, through the Kasner identities, the remaining metric exponent and dilaton magnitude. Related geometric probes give prescription-dependent constraints on terminal Kasner powers, with classical energy conditions sufficient for their late-time or finite-endpoint behavior~\cite{LiYangKasner2026}. The mass-label slope supplies the signed dilaton information directly.

Wave connection determines how this local information reaches the spectrum. At criticality, the complex endpoint $H=0$ yields a different hierarchy from the label-dependent real-tip powers: the common limit $W_\ell\to-1$ produces a shared $z^{-7/5}$ mass correction, while regular logarithms, constants and later endpoint terms retain label dependence. The gradient insertion has a different fractional power because its metric weight and dilaton profile differ. Matter sensitivity is thus distributed across response layers rather than lost. This algebraic hierarchy differs from the exponential remote contribution in the Gibbons--Maeda analysis~\cite{XiaoNianLi}. In planar Schwarzschild--AdS$_5$, the large-real-frequency scalar retarded correlator has a singularity-sensitive reflected contribution with an $\omega^{-4/3}$ correction multiplying an exponential envelope~\cite{AfkhamiJeddiEtAl2026}. Here the algebraic terms appear directly in pole responses. The five timelike examples likewise separate regular propagation from complex-endpoint scattering, with radial scaling related through the EMD geometry.

\begin{table}[!htb]
\centering\small
\renewcommand{\arraystretch}{1.2}
\begin{tabular}{@{}>{\raggedright\arraybackslash}p{.23\textwidth}
 >{\raggedright\arraybackslash}p{.35\textwidth}
 >{\raggedright\arraybackslash}p{.36\textwidth}@{}}
\toprule
Selected connection & Key response structure & Physical information and scope\\
\midrule
Spacelike real endpoint & $z^{-q}$ momentum and $z^{-\rho_\ell}$ mass terms; logarithmic resonance at $q=1$
 & Independent $p_t,p_x,p_\phi$ from momentum and the resolved label slope and intercept, Eq.~\eqref{eq:inverse}\\
\addlinespace[3pt]
Critical complex endpoint & Mass: $(B_i\Log z+C_i)/z+A_mz^{-7/5}$; gradient: $-2/(\pi z)+A_gz^{-11/10}$
 & Shared first fractional mass term; label dependence in regular and later terms\\
\addlinespace[3pt]
Fixed-label local continuation & $R_\ell=O(\log n/n)$, $R_g=O(n^{-1})$
 & Complex-endpoint mechanism across the causal threshold, in a label-dependent neighbourhood\\
\addlinespace[3pt]
Five timelike connections & $R_\ell=(B_\ell\Log z+C_\ell)/z+A_cz^{-\rho_c}+o(z^{-\rho_c})$
 & Fixed-coefficient tests at known backgrounds; radial interpretation through the EMD model\\
\bottomrule
\end{tabular}
\caption{Response structures and interior information for the selected connections. The critical powers refer to $a=1/3$; local continuation keeps the physical probe labels fixed across a neighbourhood of the causal threshold.}\label{tab:information}
\end{table}

\FloatBarrier
With physical probe labels fixed, the critical spectral family and its complex-endpoint connection continue through a local neighbourhood of the causal threshold. The real interior changes from spacelike to timelike while the response mechanism remains continuous; it is already present at criticality without an inner horizon. This variation of the background coupling differs from the overtone-resolved Kasner transitions and quantum-BTZ crossovers~\cite{HartnollZhiboedov,CovielloEtAlLong}. The comparison shows why causal character alone does not determine spectral encoding. Local scaling fixes possible powers, probe couplings distinguish metric and matter contributions, and wave connection selects which endpoint contributions enter the exterior poles (table~\ref{tab:information}).

The results combine spacelike reconstruction, fixed-label local continuation near criticality and coefficient comparisons at five known timelike backgrounds in this EMD family. An independently extracted complex-endpoint exponent would turn the EMD relation into a radial inverse map. One additional label, the low-charge reconstruction and later terms in the critical mass difference remain unresolved in the present numerical data. The main physical lesson is that high-overtone spectroscopy encodes more than local geometric powers: probe coupling supplies sensitivity to the supporting matter, and wave connection determines how both appear in the exterior spectrum.

\FloatBarrier
\Needspace{10\baselineskip}
\acknowledgments
This work was supported by the National Natural Science Foundation of China under Grant Nos.~12305063, 12547143, and 12375047, by the Hunan Provincial Natural Science Foundation of China under Grant No.~2026JJ30135, and by the Scientific Research Fund of Hunan Provincial Education Department under Grant No.~25B0823.

\appendix

\section{Source variation and connection calculations}\label{app:matching}
\subsection{Complete source and control of the simple roots}\label{app:source}
In an analytic radial coordinate $u$, write the Sturm flux state as $\bm y=(\mathcal U,\mathfrak p\mathcal U_u)^T$ and $\bm y_u=\mathsf A\bm y$. Let $\mathcal U_s,\mathcal U_{\rm UV}$ be the unit-source and unit-normalizable UV solutions and $\mathcal U_{\rm in}$ the unit-ingoing solution. With both endpoint solutions continued along the same contour, the conserved Wronskian extracts the source of the ingoing solution~\cite{MaassenVanDenBrink2004},
\begin{equation}\label{eq:generalsourcewronskian}
 C_{\rm UV}=\mathfrak p\operatorname{Wr}_u(\mathcal U_s,\mathcal U_{\rm UV})\ne0,\qquad
 \mathcal E=\frac{\mathfrak p}{C_{\rm UV}}
                 \operatorname{Wr}_u(\mathcal U_{\rm in},\mathcal U_{\rm UV}).
\end{equation}
On finite interfaces $a,b$ beyond the endpoint caps, an analytic nonzero normalization $\mathcal D=N\mathcal E$ for this contour has the ordered form
\begin{equation}\label{eq:generalorderedsource}
 \begin{aligned}
 \mathcal D&=\bm\lambda_b\mathsf T_{b\leftarrow a}\bm h_a,\\
 \bm\lambda_b&=\frac{N}{C_{\rm UV}}(\mathfrak p\mathcal U_{{\rm UV},u},-\mathcal U_{\rm UV})_b,
 &\bm h_a&=(\mathcal U_{\rm in},\mathfrak p\mathcal U_{{\rm in},u})_a^T .
 \end{aligned}
\end{equation}
For any probe strength $\xi$, its complete variation is
\begin{equation}\label{eq:compactfullvariation}
 \mathcal D_\xi=
 \bm\lambda_{b,\xi}\mathsf T_{b\leftarrow a}\bm h_a
 +\bm\lambda_b\mathsf T_{b\leftarrow a}\bm h_{a,\xi}
 +\int_a^b\bm\lambda_b\mathsf T_{b\leftarrow u}
 \mathsf A_\xi(u)\mathsf T_{u\leftarrow a}\bm h_a\,\dd u .
\end{equation}
The first two terms vary the endpoint solutions, while the integral inserts the perturbed operator along the whole connection. Changes of intermediate basis and artificial interface cancel between these terms. A common amplitude derivative multiplies $\mathcal E$ and vanishes at a root. This is why the physical response includes endpoint jets even when their indices are strength independent.

The source approximation relates the asymptotic zeros to the exact poles. On fixed nested frequency disks in the nonresonant ingoing domain, let $\mathcal D_0$ have isolated simple zeros $\omega_n^{(0)}$, a positive boundary margin and a nonzero derivative margin. If the analytic error $\mathcal D-\mathcal D_0$ is uniformly $O(n^{-\gamma_*})$ on the larger disks, Rouch\'e's theorem gives one simple zero in each smaller disk for sufficiently large $n$. The local linear bound gives $\omega_n-\omega_n^{(0)}=O(n^{-\gamma_*})$. Cauchy's estimate across the fixed disk gap, followed by Taylor expansion at the displaced root, gives
\begin{equation}\label{eq:generalactualrootcontrol}
 \mathcal D_\omega(\omega_n,0)
 =\mathcal D_{0,\omega}(\omega_n^{(0)},0)+O(n^{-\gamma_*}),
 \qquad |\mathcal D_\omega(\omega_n,0)|\ge c>0 .
\end{equation}
Evaluating $-\mathcal D_\xi/\mathcal D_\omega$ there converts the complete source derivative into a pole response. This argument applies to the spacelike, critical and five timelike connections. Uniform conclusions require common disks and positive margins. Absolute derivative bounds depend on the normalization of $\mathcal D$; the zeros and response quotient are invariant under analytic nonzero rescaling.

At criticality, on the physical exterior contour, the unit-normalizable UV solution $\mathcal U_{\rm UV}$ and unit-ingoing solution $\mathcal U_{\rm in}$ give the exterior source
\begin{equation}\label{eq:criticalexteriorsource}
 \mathcal E=\frac{P_H}{3}\operatorname{Wr}_H(\mathcal U_{\rm in},\mathcal U_{\rm UV}).
\end{equation}
Here $P_H=\mathcal G/(H-1)^2$ and $P_H\operatorname{Wr}_H(\mathcal U_s,\mathcal U_{\rm UV})=3$, so this is the same unit-endpoint source as in section~\ref{sec:probe}. For the complex-endpoint family a convenient normalization is
\begin{equation}\label{eq:canonicalnormalizedsource}
 \mathcal D=\frac{\operatorname{Wr}_X(-\omega^2\Psi_{\rm UV}/3,
                           \Psi_{\rm in}/\kappa_H)}
             {\ii\omega e^{-\ii\omega x_c}} ,
\end{equation}
Here $\Psi=S\mathcal U$, $y=x_c-X$, and
$\lim_{y\to0}\Psi_{\rm UV}/y^2=1$.
The factor $\kappa_H$ converts the continued local ingoing solution to the
chosen Jost normalization.  Both solutions span the same one-dimensional
Frobenius line, so this factor is analytic and nonzero away from its
resonances.  For the continued unit-endpoint solutions,
$\operatorname{Wr}_X(\Psi_{\rm UV},\Psi_{\rm in})
=3\mathcal E^{\rm cont}$.  Combining this identity with the framed path
comparison gives
\begin{equation}\label{eq:canonicalexternalnormalization}
 \mathcal D=\mathcal N\mathcal E^{\rm ext},\qquad
 \mathcal N=\frac{\ii\omega e^{\ii\omega x_c}}{\kappa_H}M_H\ne0.
\end{equation}
The frequency disks exclude $\omega=0$ and endpoint resonances. Because the chosen profiles preserve both endpoint indices, the endpoint solutions and these normalization factors are also analytic for sufficiently small probe strengths.
This identity relates the exact continued solutions to the exterior
source; the large-frequency Bessel expansion is applied afterwards.

At criticality, a mass variation inserts $J_m=H^{3/2}(H^{b_\ell}-1)/(H-1)^4$ into the Sturm generator; a gradient variation inserts $J_g=H^{b_{\ell_g}}/(H-1)^2$. For the continued unit-endpoint source $\mathcal E^{\rm cont}$, the bulk contraction is $-J_\xi\mathcal U_{\rm in}\mathcal U_{\rm UV}/3$, without complex conjugation. On this contour the factor in~\eqref{eq:generalorderedsource} is $N=\ii\omega e^{\ii\omega x_c}/\kappa_H=\mathcal N/M_H$. This same $N$ multiplies the complete source variation; the additional $M_H$ enters only when comparing with the exterior source.

For the complex-endpoint family the enlarged frequency disks can be chosen as
\begin{equation}\label{eq:frequencydiskcondition}
 \omega x_c/\pi=n+\beta_c+\sigma,\qquad
 |\sigma|<R<\min(1/2,\Im\beta_c).
\end{equation}
On every finite optical arm $Z=\pm t\omega x_c$, this keeps $|\Im Z|$ uniformly bounded. On the horizon tail, $X=-tx_c$ and $X_1=-t_1x_c$ with $t_1\ge t$, the ingoing Jost kernel is $[e^{2\ii\omega(X-X_1)}-1]/(2\ii\omega)$; its modulus is at most $1/|\omega|$ because $\Im(\omega x_c)>0$. The cap margin $\Re(x_c/\mathcal R_h)>0$ gives exponential decay of the potential and insertions there. These conditions hold uniformly on the stated compact near-critical sets, after reducing $R$ if needed.

The endpoint estimates retain both Bessel columns. An exponent-$s$ optical remainder gives a scaled kernel bounded near zero by
\begin{equation}\label{eq:regionalbounds}
 \|K_s(Z)\|\le C|\omega|^{-s}|Z|^{s-1-2\nu},
 \qquad Z=\omega X .
\end{equation}
Thus $s>2\nu$ makes the worst $Y_\nu^2$ entry integrable. On the large-$Z$ arms the bound is $C|\omega|^{-s}|Z|^{s-2}$: for $s>1$, integration to a fixed optical interface is $O(|\omega|^{-1})$. At the critical and timelike endpoints the geometric exponent is $2(\rho_c-1)<1$, while the probe exponents exceed one. The regular UV Green kernel supplies $O(\log|\omega|/|\omega|)$ for a mass Coulomb tail and $O(|\omega|^{-1})$ for bounded insertions. The logarithmic horizon inverse and its ingoing cap margin give an integrable Jost tail. These bounds apply to the complete first variation. At a fixed local matching point $X_0=t_0x_c$, the frequency disks keep $|\Im(\omega X_0)|$ bounded, as required for~\eqref{eq:localinsertionfrequencysplit}. Including the UV and horizon variations gives, at criticality,
\begin{equation}\label{eq:criticalcompletejetbounds}
 \mathcal D-\mathcal D_0=O(|\omega|^{-4/5}),\qquad
 \mathcal D_{\eta_{\ell_i}}=O(\log|\omega|/|\omega|),\qquad
 \mathcal D_\zeta=O(|\omega|^{-1}).
\end{equation}
The remainder is analytic on the enlarged nonresonant frequency disks. Applying~\eqref{eq:generalactualrootcontrol} with $\gamma_*=4/5$ gives actual simple poles and
\begin{equation}\label{eq:criticalactualdenominator}
 \mathcal D_\omega(\omega_n)=2\ii x_c+O(n^{-4/5}),\qquad
 |\mathcal D_\omega(\omega_n)|\ge|x_c|.
\end{equation}
These estimates hold for sufficiently large $n$. The source normalization in~\eqref{eq:canonicalnormalizedsource} is analytic and nonzero there; its derivative multiplies the vanishing exterior source and cancels from $-\mathcal D_\xi/\mathcal D_\omega$. The resulting bounds are $R_{\ell_i}=O(\log n/n)$ and $R_g=O(n^{-1})$, which justify~\eqref{eq:criticalrepresentativeabsence} independently of the explicit fractional coefficients.

For the spacelike $\nu=0$ tip the small-$Z$ bound instead contains $|Z|^{\rho-1}(1+|\log Z|)^2$, and $0<\rho<1$ also makes the remote-arm tail integrable. At $a=3$ the first geometric remainder has exponent $2/3$, giving~\eqref{eq:spacelikecompletejets}. Mass Coulomb tails account for its logarithmic derivative remainder.

\subsection{Optical connections and their geometric inputs}\label{app:connections}
We construct the optical connections at $Q=\rh=1$ using the complex-radial method of section~\ref{sec:waveconnection}. The covers and branches are specified below. Write
$\kappa=2/(1+a)=m_{\rm cov}/n_{\rm cov}$ in lowest terms.  On the spacelike cover
$v=H^{-1/n_{\rm cov}}=[r/(r+1)]^{1/n_{\rm cov}}$,
\begin{equation}\label{eq:spacelikecovers}
 P(v)=v^{3n_{\rm cov}-2m_{\rm cov}}-2^{2\kappa}(1-v^{n_{\rm cov}})^3,
 \qquad \frac{\dd x}{\dd v}=
 \frac{n_{\rm cov}v^{2n_{\rm cov}-m_{\rm cov}-1}}{P(v)}.
\end{equation}
The real tip, physical UV end and outer horizon are $v=0,1,2^{-1/n_{\rm cov}}$.
The UV action $x_p$ is obtained by continuing the residue logarithms from
$x(0)=0$ below the outer horizon.  The selected arms have
$x=tx_p$ and $x=-tx_p$, joined by a $+\pi$ optical half-turn on the same
tip inverse branch.  Their finite comparison closure makes one clockwise
outer-horizon turn and no turn about another puncture.  The same
retarded ingoing line is retained after its nonzero monodromy amplitude
is removed.  At $a=1$, for example, this prescription gives
\begin{equation}\label{eq:delta1action}
 x_p=\frac{\log2}{4}
 +\frac{3}{4\sqrt7}\arctan\frac{\sqrt7}{3}
 +\frac{\ii\pi}{4}.
\end{equation}

The critical case has $(m_{\rm cov},n_{\rm cov})=(3,2)$ and cover $v=\sqrt H$, as used to display the critical connection
in figure~\ref{fig:criticaloptical}.
For the five timelike cases instead take $v=H^{1/n_{\rm cov}}$.  The same
background gives
\begin{equation}\label{eq:timelikecovers}
 G(v)=v^{2m_{\rm cov}}-2^{2\kappa}(v^{n_{\rm cov}}-1)^3,
 \qquad \frac{\dd X}{\dd v}=-\frac{n_{\rm cov}v^{m_{\rm cov}+n_{\rm cov}-1}}{G(v)}.
\end{equation}
On this cover the complex endpoint, UV and outer horizon are $v=0,1,2^{1/n_{\rm cov}}$;
the real tip is at infinity.  The remote Bessel index is
$\nu_c=1/[2(\kappa+1)]$.
Let $h_j$ be the simple zeros of $G$ and
$\mathcal R_j=-n_{\rm cov}h_j^{m_{\rm cov}+n_{\rm cov}-1}/G'(h_j)$.  The reference action and selected
action are
\begin{equation}\label{eq:timelikereferenceaction}
 x_{c,\rm ref}=\sum_j\mathcal R_j\Log(1-h_j^{-1}),
 \qquad x_c=x_{c,\rm ref}-2\pi\ii\mathcal R_+,
\end{equation}
where the reference logarithms are principal at $v=1$ and $h_+$ is the
upper-half-plane pole nearest the physical UV end.  The local coordinate
is anchored by $X(0)=0$, $X(1)=x_c$ and $\left.\dd X/\dd H\right|_{H=1}=-1$ on the principal
UV $H$ sheet, and then continued along the two inverse lifts.

For each of the five timelike covers, numerical integration first gives
candidate inverse paths.  These candidates are enclosed by overlapping
complex disks: outward-rounded Rouch\'e inequalities give a unique optical
inverse in each disk, overlap conditions join neighbouring inverses, and
residue-log branch locks identify them with the endpoint germ.  At the UV,
the last disk overlaps the independently fixed reference inverse with
$X(1)=x_c$ and $X_H'(1)=-1$.  At the horizon $v_h=2^{1/n_{\rm cov}}$, write
$X=\mathcal R_h\Log(v-v_h)+g(v)$ and use
$w=(v-v_h)e^{g/\mathcal R_h}$; the strict cap margin
$\Re(x_c/\mathcal R_h)>0$ then completes the infinite ingoing tail.

The comparison with the real exterior also retains path order.
Truncate the UV and horizon caps, join the two tip rays by their specified
half-turn, and close the resulting path along the physical real interval.
The enclosing disks avoid all optical poles, $v=0$, and every UV image
$v^{n_{\rm cov}}=1$.  They deform the finite connection to a polygon
without crossing these punctures.  Disjoint parallel cuts from the
punctures record an ordered crossing word; interval signs fix crossing
directions and their order.  Cancelling adjacent inverse letters leaves
two clockwise outer-horizon loops at each of the five backgrounds.
The real segment from the polygon base point to the horizon crosses none
of these cuts.  Hence its surviving generator is based on the physical
ingoing line, rather than a conjugate associated with a different
continuation.  This construction supplies the source equivalence used in
section~\ref{sec:timelike}; total winding numbers alone would not suffice.

For example, at $a=1/5$ the cover is $v=H^{1/3}$ and
\begin{equation}\label{eq:timelikerepresentativeconnection}
 G(v)=v^{10}-2^{10/3}(v^3-1)^3,\qquad
 \frac{\dd X}{\dd v}=-\frac{3v^7}{G(v)},\qquad
 x_c\simeq0.1735366+0.5286460\ii.
\end{equation}
The selected complex pole is at
$v\simeq0.9128067+0.0913561\ii$ and the outer horizon is
$v_h=2^{1/3}$.  The UV and horizon inverses are joined separately to
their caps; the ordered comparison word reduces to $h_{\rm out}^{-2}$, where $h_{\rm out}$ denotes a counterclockwise loop about the outer horizon based on the physical exterior segment.
The isolated horizon cap obeys $\Re(x_c/\mathcal R_h)>0.7345$,
so its tail converges to this outer horizon.  The other four couplings
are treated separately by the same numerical construction with interval
control of the leading optical geometry.  Wave transport and the
differentiated source estimates then follow from the analytic arguments
in appendices~\ref{app:source} and~\ref{app:coefficients}.

The order-zero tip basis uses $Z=\omega x$, with the square root and logarithm continued on the selected optical branch:
\begin{equation}\label{eq:spacelikebesselbasis}
 f_0=\sqrt{\pi Z/2}J_0(Z),\qquad
 g_0=\sqrt{\pi Z/2}Y_0(Z),\qquad
 \operatorname{Wr}_Z(f_0,g_0)=1.
\end{equation}
The UV column represents the leading Bessel coefficients of $\Psi_{\rm UV}=S\mathcal U_{\rm UV}$, where $\mathcal U_{\rm UV}\sim r^{-3}$; the source branch is absent and the overall amplitude is suppressed. The $+\pi$ optical half-turn $T_0$ includes the off-diagonal term from logarithmic continuation of $Y_0$. On the companion arm the continued coefficients are expressed in a Bessel basis with argument $-\omega x$. In that basis, $(1,\ii)$ extracts the outgoing $e^{+\ii\omega x}$ coefficient up to a common nonzero factor. With $\alpha=\omega x_p-\pi/4$, the endpoint data are
\begin{equation}\label{eq:spacelikeendpointdata}
 \bm v_{\rm UV}(\alpha)=\binom{\cos\alpha}{\sin\alpha},\qquad
 \bm u_0=(1,\ii)T_0.
\end{equation}
The leading contraction $\bm u_0\bm v_{\rm UV}=\mathcal D_{0,\rm sp}(\alpha)$ is the source in~\eqref{eq:spacelikeleadingsource}; its vanishing removes the outgoing component and leaves the ingoing line.

The spacelike moment follows from the Weber--Schafheitlin identity in Ref.~\cite{NISTDLMF}, section 10.22(iv), Eq.~(10.22.57), with unit argument scale, Bessel orders $\mu=\nu=0$ and $\lambda=1-\rho$:
\begin{equation}\label{eq:besselintegral}
 I(\rho)=\int_0^\infty Z^{\rho-1}J_0(Z)^2\,\dd Z
 =\frac{2^{\rho-1}\Gamma(1-\rho)\Gamma(\rho/2)}
        {\Gamma(1-\rho/2)^3}>0,\qquad 0<\rho<1.
\end{equation}
The convergence conditions of this identity reduce to $0<\rho<1$. Expressing $Y_\nu$ through $J_{\pm\nu}$ and taking $\nu\to0$ gives the mixed moments \mbox{$\int Z^{\rho-1}J_0Y_0=-c_\rho I(\rho)$} and \mbox{$\int Z^{\rho-1}Y_0^2=(1+2c_\rho^2)I(\rho)$}. The Bessel continuation formulas in Ref.~\cite{NISTDLMF}, section 10.11, Eqs.~(10.11.1), (10.11.2) and (10.11.6), with $m=1$ and the factor $\ii$ from $\sqrt Z$, fix the half-turn $T_0$. These moments then determine $M_\rho$:
\begin{equation}\label{eq:spacelikehalfturn}
 T_0=
 \begin{pmatrix}\ii&-2\\0&\ii\end{pmatrix}.
\end{equation}
\begin{equation}\label{eq:halfcompleteborn}
 M_\rho=\frac{\pi I(\rho)}2
 \begin{pmatrix}-c_\rho&1+2c_\rho^2\\-1&c_\rho\end{pmatrix},
 \qquad c_\rho=\cot\frac{\pi\rho}{2}.
\end{equation}
The insertion on the two arms, in physical order, is
\begin{equation}\label{eq:spacelikeorderedinsertion}
 B_\rho=-M_\rho+e^{\ii\pi\rho}T_0^{-1}M_\rho T_0.
\end{equation}
At a leading zero $\alpha_0$ of $\mathcal D_{0,\rm sp}$, scale the UV column and its phase derivative by the same fixed factor $1/\cos\alpha_0$. Then
\begin{equation}\label{eq:spacelikeendpointvectors}
 \bm v_0=\binom{1}{\ii/3},\qquad
 \bm v_1=\binom{-\ii/3}{1},\qquad
 \bm u_0\bm v_1=-\frac83.
\end{equation}
This coefficient-space convention is distinct from the unit UV endpoint normalization of section~\ref{sec:probe}. Equations~\eqref{eq:spacelikehalfturn} and~\eqref{eq:halfcompleteborn} give $\bm u_0B_\rho\bm v_0=\mathcal P(\rho)$, with the nonzero projection displayed in~\eqref{eq:physicalprojectionmain}.

At $a=1/2$, the subleading momentum coefficient also requires the
finite-part continuation of the complete insertion matrix. Here the background mass parameter takes the value $M=2^{8/3}$. Define the momentum primitive $F_k=3v$, for which $V_k\dd x=3\dd v$.
The first-order Liouville coordinate
$y=x-k^2 F_k/(2\omega^2)$ moves the UV end to
$y_p=x_p-3k^2/(2\omega^2)$ and leaves the UV inverse-square index
unchanged.  Its residual operator is
$\mathcal Q=V_gV_k+F_kV_g'/2-V_k''/4
 =cy^{-5/2}/16+O(y^{-2})$, with $c=-\ii\sqrt{3M/2}$.
The transformed insertion has overall factor $(-c/16)k^2\omega^{-3/2}$
and Bessel exponent $-1/2$.  Integrating its complete matrix on a fixed
unit optical half-circle is absolutely convergent on both infinite
arms; that half-circle is not shrunk to the origin.  It is holomorphic
for $\Re s<1$, agrees with the ordered matrix $B_s$ in
\eqref{eq:spacelikeorderedinsertion}, formed from $M_s$ in
\eqref{eq:halfcompleteborn}, for $0<\Re s<1$, and hence extends it by
uniqueness.  The large-arm-mean finite part at $s=3/2$ obeys
$B_{-1/2}^{(1)}=16B_{3/2}^{\rm FP}$ and
$I(-1/2)=16I(3/2)$.  Its nonzero physical contraction is
\begin{equation}\label{eq:halfmatrixcontinuation}
 (\ii,-3)B_{3/2}^{\rm FP}\binom{1}{\ii/3}
   =-\frac{2\pi I(3/2)}3(1+\ii)\ne0.
\end{equation}
In the complete source derivatives the moved UV phase and the first
$\omega^{-1/2}$ geometric correction occur in the same phase-derivative
factor $\mathcal J$, obtained by differentiating the leading source
together with its first $\omega^{-1/2}$ geometric correction with respect
to $\alpha$.  Thus
\begin{equation}\label{eq:halfsourcejets}
\begin{aligned}
 \mathcal D_\omega&=x_p\mathcal J+O(\Log\omega/\omega),\\
 \mathcal D_{k^2}&=-3\mathcal J/(2\omega)
 +(-c/16)\omega^{-3/2}G_{-1/2}
 +O(\Log\omega/\omega^2).
\end{aligned}
\end{equation}
Here $G_{-1/2}$ contracts $B_{-1/2}^{(1)}$ with the same endpoint data;
at the leading phase it is sixteen times the $B_{3/2}^{\rm FP}$
contraction in~\eqref{eq:halfmatrixcontinuation}.  The UV column uses the
same fixed normalization in $\mathcal J$ and $G_{-1/2}$.
Their quotient cancels the apparent regular--geometry cross term at
$3/2$.  The remaining nonzero full-matrix projection gives
\begin{equation}\label{eq:halfmomentumcoefficient}
 \mathcal A_k=
 \frac{\pi I(3/2)\sqrt{3M/2}}{4x_p}(1-\ii)\Omega_p^{-3/2},
 \qquad
 I(3/2)=-\frac{2\sqrt{2\pi}\Gamma(3/4)}{\Gamma(1/4)^3}.
\end{equation}
All powers follow the already selected optical logarithm.  At $a=1$,
the pole of $I(\rho)$ instead joins the propagation term to give
\eqref{eq:delta1log}.

\subsection{Critical connection and local continuation}\label{app:criticalpath}
The critical optical poles and residues are
\begin{equation}\label{eq:criticalexactresidues}
 h_+=\frac{5+\ii\sqrt3}{7},\qquad
 \mathcal R_c=\frac{-\sqrt{2\sqrt7-1}+\ii\sqrt{2\sqrt7+1}}{6\sqrt{14}},
 \qquad \mathcal R_h=\frac1{3\sqrt2}.
\end{equation}
With $P=-2\pi\ii\mathcal R_c$, $I=\int_0^1H^{3/2}\mathcal G(H)^{-1}\dd H$ and $x_c=P-I$, the elementary primitive of $H/\mathcal G$ gives
\begin{equation}\label{eq:criticalactionanalytic}
 0<I<\int_0^1\frac H{\mathcal G}\,\dd H
 =\frac{\pi}{3\sqrt3}<2\pi\Im\mathcal R_c,\qquad
 \Re x_c>0,\quad 0<\Im x_c=-2\pi\Re\mathcal R_c<\pi\mathcal R_h .
\end{equation}
These signs fix the two germs, with $\theta=\arg x_c$:
\begin{equation}\label{eq:criticalpositivegerm}
 \begin{aligned}
 H_+(t)&=(20|x_c|t)^{2/5}e^{2\ii(\pi+\theta)/5}[1+O(t^{2/5})],
       &&X=tx_c,\\
 H_-(t)&=(20|x_c|t)^{2/5}e^{2\ii(2\pi+\theta)/5}[1+O(t^{2/5})],
       &&X=-tx_c .
 \end{aligned}
\end{equation}
On the upper bank of the principal cut,
\begin{equation}\label{eq:criticalcut}
 X_{\rm cut}(s)=-\ii\int_0^s
   \frac{t^{3/2}\,\dd t}{8(t+1)^3-t^3},\qquad
 X_{\rm cut}(\infty)=\ii(\Im x_c-\pi\mathcal R_h).
\end{equation}
Both values are negative imaginary for $s>0$.  The upper half-plane has
one optical pole, so $V=\Im(X/P)$ is single-valued there.  It tends to
$-\infty$ at that pole and has nonpositive limits at zero, the cut and
infinity.  On the positive germ,
$V=tI\Im x_c/|P|^2$ increases.  A small horizon cap is inaccessible to
this forward flow: in its coordinate
$w=(H-2)e^{g/\mathcal R_h}$, $\dot w=(x_c/\mathcal R_h)w$ points outward.
Outside that cap $V$ is bounded above.  Thus the positive arm reaches
the real boundary in finite time; on every interval $t\ge t_0>0$ the
excluded boundary limits leave a compact region in which the inverse
vector field is bounded.  An interior endpoint would allow continuation,
while arrival above $H=2$ is excluded by
$\Im\dot H_+=\Im x_c/X_H'>0$.

The arm has no self-intersection, since one would require
$(t_2-t_1)x_c=jP$ although $P/x_c$ is not real.  Closing it along the
real axis gives a clockwise Jordan lobe.  Its period is either zero or
$P$; at the first contact, imaginary parts force $t=1$ and period $P$.
The strictly decreasing real primitive on $(0,2)$ then fixes the contact
at $H=1$, fixing both the UV endpoint and its inverse branch.

Let $K$ be the closed disk bounded by this arm and its reflection.
It contains the two complex poles and touches the principal cut only at
zero.  The slit plane with $K$ removed is simply connected; removing
$H=2$ leaves an exterior domain whose fundamental group has only the
outer-horizon generator.  In that domain $\Re X$ is single-valued,
because the remaining periods are $2\pi\ii j\mathcal R_h$.
The two boundary branches use the primitive continued from the same real point of the physical exterior; Schwarz reflection fixes their real parts consistently.
It is nonnegative on the lobe boundary and zero on the cut and at
infinity; its only negative divergence is at the outer horizon.
At sufficiently small starting radius, the prescribed $+\pi$ companion germ lies outside $K$ and obeys
$\Re X=-t\Re x_c$.  Finite negative bands are compact, so it extends
for all finite $t$ and approaches $H=2$.  The logarithmic cap then gives
$w(t)=w(t_0)e^{-(t-t_0)x_c/\mathcal R_h}$ and controls the entire tail.

To compare wave solutions, first truncate both endpoint caps and displace
the positive arm and the half-turn into a pole-free collar outside $K$.
The exact wave equation is analytic throughout this finite comparison
domain.  The UV solution with unit $r^{-3}$ coefficient has Frobenius
index three and recursion denominators $n(n+3)$ for $n>0$, so it
continues from the arm through the UV cap to the physical interval
without acquiring a source branch.  The horizon cap retains the
ingoing exponent $\alpha=-\ii\omega\mathcal R_h$.
The comparison paths differ by an integer number $j$ of based
outer-horizon loops.  Their monodromy acts on this line as
$e^{2\pi\ii j\alpha}=e^{2\pi j\omega\mathcal R_h}$, an analytic
nonzero scalar.  The winding comparison applies to the truncated paths;
the infinite spiral is completed by its Frobenius cap.  These facts give
the exact source relation~\eqref{eq:criticalretardedequivalence} and the
normalization~\eqref{eq:canonicalexternalnormalization}.

\label{app:localcontinuation}
Put $\kappa=2/(1+a)$, so $\rho_c-1=(\kappa+1)^{-1}=2\nu_c$ and $b_\ell=\ell\sqrt{\kappa(2-\kappa)}$. The nearby optical family is
\begin{equation}\label{eq:localopticalfamily}
 G_\kappa=H^{2\kappa}-2^{2\kappa}(H-1)^3,\qquad
 X_\kappa'=-H^\kappa/G_\kappa,\qquad X_\kappa(0)=0 .
\end{equation}
Here the prime on $X_\kappa$ denotes differentiation with respect to $H$. The moving simple upper pole and the UV reference branch define
\begin{equation}\label{eq:localaction}
 x_c(\kappa)=-2\pi\ii\mathcal R_c(\kappa)-I_\kappa,\qquad
 I_\kappa=\int_0^1
 \frac{H^\kappa\,\dd H}{H^{2\kappa}+2^{2\kappa}(1-H)^3}.
\end{equation}
The upper pole remains simple near $\kappa=3/2$, the action is analytic, and its strict critical gaps persist. The truncated path and its comparison homotopy lie in a compact set where $G_{3/2}\ne0$; uniform convergence preserves pole exclusion nearby. Overlapping inverse charts identify the independent UV branch with $X_\kappa(1)=x_c$, $X_\kappa'(1)=-1$. The common horizon cap has residue $\mathcal R_h=2^{-\kappa}/(3-\kappa)$ and retains a positive $\Re(x_c/\mathcal R_h)$ margin. This continues the entire framed source comparison, including its complete negative tail. For $\kappa>3/2$, the new inner horizon escapes to $H=\infty$ as $\kappa\downarrow3/2$ and lies outside the compact comparison domain.

Keep the physical labels fixed in compact sets
\begin{equation}\label{eq:compactlabels}
 K_m\Subset(2/\sqrt3,\sqrt3),\qquad
 K_g\Subset(1/\sqrt3,2/\sqrt3).
\end{equation}
With $b_{\ell_g}=\ell_g\sqrt{\kappa(2-\kappa)}$, the common mass, label-dependent mass and gradient exponents are, respectively,
\begin{equation}\label{eq:compactendpointgrades}
 \rho_c,\qquad \rho_c+(\rho_c-1)b_\ell,\qquad
 \rho_{c,g}=(\rho_c-1)(2+b_{\ell_g}).
\end{equation}
The geometric exponent is $2(\rho_c-1)$. Fixed physical labels mean that $b_\ell$ and $b_{\ell_g}$ vary with the background through these relations. The real-tip reference exponents on the two causal sides are
\begin{subequations}\label{eq:localrealtipgrades}
\begin{align}
 \rho_m(\kappa)&=
 \begin{cases}
 (3-\kappa-b_\ell)/(2-\kappa), & \kappa<3/2\quad\text{(spacelike)},\\
 (\kappa-b_\ell)/(\kappa-1), & \kappa>3/2\quad\text{(timelike)},
 \end{cases}
 \label{eq:localrealtipgrades:mass}\\[4pt]
 \rho_g(\kappa)&=
 \begin{cases}
 (1-b_{\ell_g})/(2-\kappa), & \kappa<3/2\quad\text{(spacelike)},\\
 2-b_{\ell_g}/(\kappa-1), & \kappa>3/2\quad\text{(timelike)}.
 \end{cases}
 \label{eq:localrealtipgrades:gradient}
\end{align}
\end{subequations}
Their critical limits are $3-\sqrt3\ell$ and $2-\sqrt3\ell_g$. The strict exponent and action margins select a two-sided neighbourhood $\mathcal U(K_m,K_g)$ in which
\begin{equation}\label{eq:compactstrictmargins}
 1<b_\ell<\min(\kappa,3-\kappa),\qquad
 \kappa-1<b_{\ell_g}<\min(1,2(\kappa-1))
\end{equation}
hold with common positive gaps. They make both Bessel columns integrable, put the probe exponents above one and the real-tip exponents strictly between zero and one.

On every closed $\mathcal J\Subset\mathcal U$, the regional wave bounds and ingoing cap are uniform. The leading source $\mathcal D_0$ has uniform simple-zero boundary and derivative margins on the nested frequency disks of appendix~\ref{app:source}. The source/root argument there gives the spectrum~\eqref{eq:localcontinuationspectrum}, with $\gamma_*=\inf_{\mathcal J}2(\rho_c-1)>0$, and controls its actual-root derivative:
\begin{equation}\label{eq:compactactualdenominator}
 \begin{aligned}
 \mathcal D_\omega(\omega_n)&=2\ii x_c+O(n^{-\gamma_*}),
       &|\mathcal D_\omega(\omega_n)|&\ge c_{\mathcal J}>0 ,
 \end{aligned}
\end{equation}
and
\begin{equation}\label{eq:compactresponsebounds}
 \sup_{\mathcal J\times K_m}|R_\ell(n)|
       \le C_m\frac{\log n}{n},\qquad
 \sup_{\mathcal J\times K_g}|R_g(n;\ell_g)|\le\frac{C_g}{n}.
\end{equation}
The constants may depend on $\mathcal J$ and the fixed label sets. Since the reference exponents have a common gap from one,
\begin{equation}\label{eq:compactscaledabsence}
 \sup_{\mathcal J\times K_m}|z_n^{\rho_m}R_\ell|\longrightarrow0,\qquad
 \sup_{\mathcal J\times K_g}|z_n^{\rho_g}R_g|\longrightarrow0 ,
\end{equation}
hold uniformly on these compact sets. On the timelike side the reference uses proper-radial scaling. These estimates establish~\eqref{eq:localcontinuationspectrum} and~\eqref{eq:localcontinuationabsence} for the local continued family; they do not extend the connection to the full timelike interval.

\subsection{Fractional coefficients and restoration of the regular mean}\label{app:coefficients}
The finite part of the Bessel insertion determines the endpoint contribution, and its two-arm projection fixes the fractional response coefficient. Restoring the regular mean and UV phase then gives the complete pole response, including the correction orders used in the numerical comparisons.
For $f_\nu=\sqrt{\pi Z/2}J_\nu(Z)$ and $g_\nu=\sqrt{\pi Z/2}Y_\nu(Z)$, Eqs.~(10.11.1)--(10.11.2) of Ref.~\cite{NISTDLMF} with $m=1$ give the $+\pi$ optical half-turn
\begin{equation}\label{eq:complexhalfturn}
 T_+(\nu)=\begin{pmatrix}\ii e^{\ii\pi\nu}&-2\cos\pi\nu\\0&\ii e^{-\ii\pi\nu}\end{pmatrix}.
\end{equation}
An insertion with exponent $\rho$ uses the complete two-column kernel
\begin{equation}\label{eq:completebesselkernel}
 \mathcal R_\rho=Z^{\rho-2}
 \begin{pmatrix}f_\nu g_\nu&g_\nu^2\\-f_\nu^2&-f_\nu g_\nu\end{pmatrix},
 \qquad J=\begin{pmatrix}0&1\\-1&0\end{pmatrix}.
\end{equation}
In the first-order Bessel construction of section~\ref{sec:waveconnection}, the lower endpoint is integrable for $1<\rho<2$ and $\rho>2\nu$, while the large-$Z$ mean must be separated:
\begin{equation}\label{eq:timelikeFPdefinition}
 H_{\rm FP}=
 \lim_{\Lambda\to\infty}
 \left[\int_0^\Lambda{\cal R}_\rho(Z)\dd Z
             -\frac{\Lambda^{\rho-1}}{2(\rho-1)}J\right].
\end{equation}
The mean-zero oscillating tail converges by integration by parts; the next tail is absolutely integrable. For $2\nu<\rho<1$, the ordinary product moments follow from Eq.~(10.22.57) of Ref.~\cite{NISTDLMF}, with orders $\pm\nu$ and $\lambda=1-\rho$. The complete mean subtraction in~\eqref{eq:timelikeFPdefinition} identifies their continued Gamma expressions with $H_{\rm FP}$ for $1<\rho<2$, $\rho>2\nu$, without an arbitrary constant matrix.

The UV sheet also fixes the regular complex constant. For the mass one-form~\eqref{eq:subcriticaloneform}, $\Log(1-H)=\Log(H-1)-\ii\pi$ and its primitive has finite part $I_{\rm FP}=b_\ell-1-b_\ell[\gamma_E+\psi(b_\ell)]+\ii\pi b_\ell$. Joining this finite arm phase to the regular UV Coulomb solution gives the frequency-dependent logarithm and constant below. The standard Coulomb phase and large-argument expansion are Eqs.~(33.2.9)--(33.2.10) and (33.11.1) of Ref.~\cite{NISTDLMF}, with angular index $L_C=1$ and, to first order in the probe strength, Coulomb parameter $q_C=\eta_\ell b_\ell/(2\omega)$ in the scaled UV coordinate $\omega y$. The solution is continued on the UV sheet fixed above, giving
\begin{equation}\label{eq:uvregularphasematching}
 I_{\rm FP}+b_\ell[\Log(2\omega)-\psi(2)]
 =b_\ell[\Log(-2\omega)-\psi(b_\ell)]-1=F(\omega).
\end{equation}
Here $\psi(2)=1-\gamma_E$ and the selected fourth-quadrant branch obeys $\Log(-2\omega)=\Log(2\omega)+\ii\pi$. The critical mass coefficients~\eqref{eq:criticallabelregular} use the same matching for each $b_\ell$.

For $\rho=1+2\nu$ the endpoint finite part is
\begin{equation}\label{eq:subcriticalmatrixfinitepart}
\begin{aligned}
 H_{\rm FP}&=h
 \begin{pmatrix}\tan\pi\nu&1\\-1&-\tan\pi\nu\end{pmatrix},\\[3pt]
 h&=\frac{\pi2^{2\nu-1}\Gamma(-2\nu)\Gamma(1/2+2\nu)}
 {\sqrt\pi\,\Gamma(1/2-\nu)^2}.
 \end{aligned}
\end{equation}
For the timelike mass insertion, set $\nu=\nu_c$ and $\rho_c=1+2\nu_c$. Since $1<\rho_c<2$ and $\rho_c>2\nu_c$, the mean/finite-part decomposition~\eqref{eq:localinsertionfrequencysplit} applies. Its ordered two-arm insertion is
\begin{equation}\label{eq:timelikeorderedinsertion}
 C_1=e^{\ii\pi\rho_c}H_{\rm FP}T_+-T_+H_{\rm FP}.
\end{equation}
For the UV vector $r_{\rm UV}=(\cos A_\omega,\sin A_\omega)^T$, $A_\omega=\omega x_c-\pi\nu_c/2-\pi/4$, and outgoing row $\lambda_{\rm out}=(1,\ii)$, the leading-root contraction is
\begin{equation}\label{eq:subcriticalglobalcontraction}
 \mathcal Q_c=
 \frac{\lambda_{\rm out}C_1r_{\rm UV}}{\mathfrak a_\omega}
 =2\ii h e^{\ii\pi\nu}
       \frac{\cos(2\pi\nu)}{\cos(\pi\nu)}\ne0 .
\end{equation}
Here $\mathfrak a_\omega=\ii e^{-\ii(A_\omega+\pi\nu_c)}$. The Gamma factor $h$ is finite and negative for the timelike range, while both cosine factors are positive. The endpoint insertion therefore survives the physical two-arm projection.

The leading-root identities $\lambda_{\rm out}J=-\ii\lambda_{\rm out}$ and $\lambda_{\rm out}T_+r_{\rm UV}=0$ exhibit the common horizon amplitude; its derivative drops out at an actual zero of the complete source. The UV-side mean restores the short-arm phase; joining it to the remaining arm and regular Coulomb solution gives $F(\omega)$ in~\eqref{eq:uvregularphasematching}. Evaluated at the actual simple roots, the complete derivative combines that restored mean with the fractional contraction:
\begin{equation}\label{eq:subcriticalgeneralrootjets}
 \mathcal D_{\eta_\ell}=-\ii F(\omega)/\omega
       -\alpha_c\mathcal Q_c\omega^{-\rho_c}
       +O_a(|\omega|^{1-2\rho_c}\log|\omega|).
\end{equation}
The frequency derivative from appendix~\ref{app:source} is $\mathcal D_\omega(\omega_n)=2\ii x_c+O_a(n^{-2(\rho_c-1)})$. Thus $R_\ell=-\mathcal D_{\eta_\ell}/\mathcal D_\omega$ yields~\eqref{eq:subcriticalcomplete}, with $A_c=\alpha_c\mathcal Q_c/(2\ii x_c\Omega_c^{\rho_c})$ and the stated remainder. The physical coefficient has the finite nonzero Gamma prefactor
\begin{equation}\label{eq:timelikeGammaprefactor}
 \mathcal F(\epsilon)=
 \frac{2^{1-\epsilon}\epsilon^{1-\epsilon}\Gamma(-\epsilon)}
 {\sqrt\pi\,\Gamma((1-\epsilon)/2)^2}
 \frac{\Gamma(1/2+\epsilon)\cos(\pi\epsilon)}
 {\cos(\pi\epsilon/2)}.
\end{equation}
Here $\epsilon$ is the argument of $\mathcal F$; the physical coefficient uses $\epsilon=\rho_c-1=2\nu_c$. This fixes the amplitude and phase before the numerical comparison.

At criticality, a local term $c_{\rm loc}(-20X)^{\rho-2}$ gives
\begin{subequations}\label{eq:criticalcompletecoefficient}
\begin{align}
 \mathcal K(\rho)&=\frac{\ii\pi^2e^{\ii\pi \rho/2}\Gamma(1-\rho)}
 {2^{2-\rho}\cos(\pi/5)\Gamma(1-\rho/2)^2
  \Gamma(6/5-\rho/2)\Gamma(4/5-\rho/2)},\\
 A(\rho,c_{\rm loc})&=-\frac{c_{\rm loc}20^{\rho-2}e^{\ii\pi(\rho-2)}\mathcal K(\rho)}
                    {x_c\Omega_c^\rho}.
\end{align}
\end{subequations}
Here we use $\Log(-20)=\log20+\ii\pi$ and continue all frequency powers on the same logarithmic branch. This coefficient includes the full-mean subtraction and physical two-arm contraction. For a fixed critical gradient label, the restored positive-arm mean is $-\int_0^1H^{b_{\ell_g}-3/2}\dd H=-1/(b_{\ell_g}-1/2)$. For $b_{\ell_g}=3/4$ it gives the regular coefficient $-2/\pi$ in~\eqref{eq:criticalsurvivingresponse}.

Separating the complete source into its full-arm phase, nonzero analytic
amplitude $\mathscr A$ and remote connection ratio $\mathscr C$ gives
\begin{equation}\label{eq:criticalphasequotientsource}
 \begin{aligned}
 \mathcal D&=\mathscr A\left[
 \frac{e^{2\ii[\omega x_c-\Phi(\omega,\xi)]}}
      {\mathscr C(\omega,\xi)}-1\right],\\
 e^{2\ii[\omega x_c-\Phi]}&=\mathscr C
 \quad\hbox{at a pole}.
 \end{aligned}
\end{equation}
The phase $\Phi$ contains the restored arm mean and the regular UV
Coulomb phase.  Differentiating at zero insertion strength removes the
common amplitude, since both of its derivatives multiply the vanishing
bracket.  The response is therefore
\begin{equation}\label{eq:criticalphasequotientresponse}\begin{aligned}
 R_\xi&=
 \frac{2\ii\Phi_\xi+(\Log\mathscr C)_\xi}
 {2\ii x_c-2\ii\Phi_{\omega,0}-(\Log\mathscr C_0)_\omega},
 \\
 \Phi_\xi&=\frac{F_\xi(\omega)}{2\omega}+O(|\omega|^{-3}\log^2|\omega|),
 \qquad \Phi_{\omega,0}=O(|\omega|^{-2}).
\end{aligned}
\end{equation}
The subscript $0$ denotes zero insertion strength.  $F_\xi$ denotes
the regular coefficient of the selected strength
direction: $F_{\rm mass}=b_\ell[\Log(-2\omega)-\psi(b_\ell)]-1$
and $F_{\rm gradient}=-4$.  For the critical endpoint, $C_c=\ii/[2\cos(\pi/5)]$. The first correction to the denominator
is found by retaining the baseline connection through the exponents
$4/5,6/5,8/5$:
\begin{equation}\label{eq:criticalbaselinegrades}
\begin{aligned}
 \Log\mathscr C_0&=\Log C_c+c_{4/5}\omega^{-4/5}
   +c_{6/5}\omega^{-6/5}+c_{8/5}\omega^{-8/5}
   +O(|\omega|^{-2}\log^2|\omega|),\\
 (\Log\mathscr C_0)_\omega&=O(|\omega|^{-9/5}).
 \end{aligned}
\end{equation}
The $8/5$ term includes two ordered geometric insertions.  Differentiating
the explicit powers and applying Cauchy's estimate on the nested disks of
appendix~\ref{app:source} gives the second line.  Combined with the
regular numerator $O(\omega^{-1}\Log\omega)$, it places the first
regular-response denominator correction at exponent $14/5$.

The retained probe insertions all have $1<\rho<2$, so the mean subtraction in
\eqref{eq:timelikeFPdefinition} suffices for their local coefficients.
There is also no independent regular $z^{-2}$ term.  The second smooth-arm
Riccati term $-V'/(4\lambda^2)$, $\lambda=\pm\ii\omega$, changes the
two waves by the same amplitude.  In the regular UV Coulomb core, Eq.~(33.11.1), with the parameters in Eq.~(33.8.3) of Ref.~\cite{NISTDLMF}, gives
$a_1(q_C)=q_C/2+\ii(2+q_C^2)/2$ for $L_C=1$. Its linear contribution is common to both waves in the matched smooth-arm basis.  The regular-column
phase moment of the constant insertion has zero finite part after its
mean and matching endpoint are removed.  The remaining UV Taylor terms
are $O(y)$, $y=x_c-X$.  Matching their parameter jets to the same
smooth-arm basis cancels the order-two interface terms and leaves
$O(\omega^{-3}\log^2|\omega|)$ in the phase response.

The first geometric--probe interaction fixes the remaining asymptotic
order.  Put $\gamma_{\rm geom}=4/5$ and
$\vartheta_\nu(Z)=Z-\pi\nu/2-\pi/4$, $\nu=1/5$.  The leading oscillatory kernel and
its bounded zero-mean primitive are
\begin{equation}\label{eq:criticaloscillatorykernel}\begin{aligned}
 \mathcal R_{\rm osc}&=\frac12
 \begin{pmatrix}\sin2\vartheta_\nu&-\cos2\vartheta_\nu\\-\cos2\vartheta_\nu&-\sin2\vartheta_\nu\end{pmatrix},
 \\
 P_0&=\frac14
 \begin{pmatrix}-\cos2\vartheta_\nu&-\sin2\vartheta_\nu\\-\sin2\vartheta_\nu&\cos2\vartheta_\nu\end{pmatrix},
\qquad P_0'=\mathcal R_{\rm osc}.
\end{aligned}
\end{equation}
At a fixed remote optical interface, the Bessel endpoint also has an
oscillatory term proportional to $\omega^{-2}P_0$, including for the
first unexpanded mass layer with $\rho=11/5>2$.  It cancels the oppositely
oriented endpoint term of the adjacent smooth-arm basis; this cancellation
is required for the mass remainder stated below.
Keeping the mean rotations as exact exponentials, the first cross
generator after the one-insertion gauges is
\begin{equation}\label{eq:criticalmixednormalformorder}
 N_{\gamma_{\rm geom} \rho}=-\frac14
 \left(\frac1{\gamma_{\rm geom}-1}+\frac1{\rho-1}\right)
 Z^{\rho+\gamma_{\rm geom}-3}[J,\mathcal R_{\rm osc}]
 +O(Z^{\rho+\gamma_{\rm geom}-4}).
\end{equation}
The displayed commutator is the derivative of $[J,P_0]$, while the
mean--mean commutator vanishes.  Removing this interface primitive
leaves an integrable kernel for $\rho+\gamma_{\rm geom}<3$, so the first connected
correction has exponent $\rho+4/5$.  The same order includes the change
of a local projection when the leading phase is replaced by the actual
root phase.  The first unexpanded common mass layer and its geometric
interaction both have exponent $11/5$; for the gradient they both have
exponent $19/10$.

Set $w=(-20X)^{1/5}$ on the selected branch. Optical reversion gives
$H=w^2-6w^4/7+O(w^6)$; with $R_{\ell_i,\rm reg}=(B_i\Log z+C_i)/z$, the retained response expansions are

\begin{subequations}\label{eq:criticalcompleteexpansions}
\begin{align}
 R_{\ell_1}={}&R_{\ell_1,\rm reg}+A(7/5,-8)z^{-7/5}
              +A(9/5,-16/7)z^{-9/5}\notag\\
           &+A(19/10,8)z^{-19/10}+O(|z|^{-11/5}\log^2|z|),\\
 R_{\ell_2}={}&R_{\ell_2,\rm reg}+A(7/5,-8)z^{-7/5}
              +A(9/5,-16/7)z^{-9/5}\notag\\
           &+A(37/20,8)z^{-37/20}+O(|z|^{-11/5}\log^2|z|),\\
 R_g={}&-2/(\pi z)+A(11/10,8)z^{-11/10}
             +A(3/2,-60/7)z^{-3/2}\notag\\
      &+O(|z|^{-19/10}\log^2|z|).
\end{align}
\end{subequations}
The remainders apply to the complete responses at the actual simple roots.

\section{Numerical method and resolution of the responses}\label{app:numerics}
\subsection{Exterior boundary-value problem and tracking}
The horizon-series approach provides an early numerical construction of AdS QNMs~\cite{HorowitzHubeny}. We use the two-endpoint Frobenius construction of appendix D of Ref.~\cite{HartnollZhiboedov}, matching an ingoing horizon series to a normalizable UV series at an interior point of the exterior radial interval. The ingoing factor is removed before the series expansion; algebraic compactifications of $r/(r+Q)$ make the selected background and probe coefficients polynomial. The matching determinant is a Wronskian source function, and its zeros impose both physical endpoint conditions. No interior wave boundary condition enters this calculation.

Differentiating both Frobenius recurrences with respect to frequency, $k^2$ and the probe strengths gives the derivatives in~\eqref{eq:evansidentity}, including changes of both endpoint series. For the gradient response, differentiation with respect to the independent parameter $\zeta$ gives the mixed derivative. Central differences at nonzero strengths provide a separate check. The recurrences use only the exterior background and probe equations; geometric Kasner values and analytic response exponents enter the subsequent comparisons.

Newton iteration uses the source and its frequency derivative. Frequencies are continued from neighbouring poles or previously checked state points; overlapping and reverse-tracked segments test branch consistency. Raising arithmetic precision and truncation, changing the matching fraction and comparing radial compactifications tests finite-setting convergence of frequencies and responses. For example, the critical $n=640$ response uses 2600-digit arithmetic and truncation 9600, checked against 3600 digits and truncation 12800 with a changed matching fraction.

For the five timelike backgrounds, the frequency branches are continued through every integer $n=241,\ldots,320$. Complete responses are recalculated at $n=240$ to check overlap and evaluated at the discrete overtones $n=280,320$, with the same probe profile. At these response points, the roots are refined using the determinant and its frequency derivative before evaluating the complete response derivatives; the $n=320$ results are checked by increasing precision and truncation and changing the matching fraction. The all-integer response sample remains $n=41,\ldots,240$.

The exterior calculation supplies frequencies and complete response derivatives independently of the optical-path construction. Agreement with the predicted spacing and fixed complex coefficients supports identification of the tracked modes with the asymptotic family over the computed windows. The optical construction establishes the path homotopy; the exterior calculation tests its spectral consequences.

\subsection{Free exponents and fixed-coefficient comparisons}
Table~\ref{tab:probelabels} gives the labels for the spacelike inverse and the corresponding analytic comparison exponents.

\begin{table}[H]
\centering\small
\renewcommand{\arraystretch}{1.12}
\begin{tabular}{@{}cccccc@{}}
\toprule
Background & \multicolumn{2}{c}{Probe labels} & \multicolumn{3}{c}{Analytic response exponents}\\
\cmidrule(lr){2-3}\cmidrule(l){4-6}
$a$ & $\ell_1$ & $\ell_2$ & $q$ & $\rho_{\ell_1}$ & $\rho_{\ell_2}$\\
\midrule
$1/2$ & $7/(4\sqrt2)$ & $9/(4\sqrt2)$ & $3/2$ & $3/4$ & $1/4$\\
$1$ & $5/4$ & $7/4$ & $1$ & $3/4$ & $1/4$\\
$3$ & $\sqrt3$ & $4/\sqrt3$ & $2/3$ & $2/3$ & $1/3$\\
$5$ & $\sqrt5$ & $7/\sqrt5$ & $3/5$ & $3/5$ & $1/5$\\
$7$ & $\sqrt7$ & $10/\sqrt7$ & $4/7$ & $4/7$ & $1/7$\\
\bottomrule
\end{tabular}
\caption{Probe labels and analytic reference exponents for the spacelike reconstruction in table~\ref{tab:spacelikegrades}. The QNM response exponents are extracted independently.}\label{tab:probelabels}
\end{table}

For $a=1/2$, the selected mass fits retain freely fitted corrections at offsets $1,3/2$ and an independent regular $1/z$ term after subtraction of the known UV logarithm; templates containing a half-step correction enter the sensitivity comparison. Momentum is fitted after the subtraction in~\eqref{eq:halfresponsesmain}. At $a=1$ the mass-label fits retain two integer corrections and the momentum fit uses the resonant basis with a regular term and allowed logarithmic corrections. The selected $a=3,5,7$ templates include the regular propagation and local fractional corrections, and are compared with UV-log-subtracted and alternative remainder templates. 

For spacelike reconstruction, the real leading exponents are fitted freely and the complex amplitudes are determined by relative least squares. We compare local minima and vary the frequency window and remainder model before inserting the exponents into~\eqref{eq:inverse}. The metric and dilaton exponents from the background, together with the Kasner relations, are then used to assess the reconstruction. The selected all-integer extraction windows are $n=180,\ldots,300$ for $a=1/2$, $240,\ldots,300$ for $a=1$, and $220,\ldots,300$ for $a=3,5,7$.

At the momentum resonance $a=1$, the smooth basis
\begin{equation}\label{eq:marginalbasis}
 f_q(z)=\begin{cases}
 (z^{-q}-z^{-1})/(1-q),&q\ne1,\\
 (\Log z)/z,&q=1.
 \end{cases}
\end{equation}
is used alongside an independent regular term. In table~\ref{tab:spacelikegrades}, the $a=1/2$ row is rounded to three decimal places because its remainder-model variation is of order $10^{-3}$, larger than the fit residual. The other rows use four decimal places, reflecting their sensitivity to the window and remainder model. The additional $a=7$, $\ell=5/\sqrt7$ response admits fits close to its predicted exponent, but competing remainder templates do not give a stable extraction. At $a=5$, $Q/\rh=2/5$, the roots and responses converge and the selected momentum fit is close to the geometric prediction, but an alternative remainder template admits a competing exponent. This state is therefore excluded from the twelve-state reconstruction; the issue concerns extraction from a weak response, not the existence or numerical convergence of its poles.

For the critical and timelike comparisons, the analytic responses and displayed ratios are evaluated at each numerical complex frequency with the analytic exponents and complex coefficients fixed. The complex relative RMS error is
\begin{equation}\label{eq:numericalrms}
 \left[\frac1N\sum_{n\in\mathcal W}
 \left|\frac{R_{\rm num}(n)-R_{\rm pred}(n)}{R_{\rm num}(n)}\right|^2
 \right]^{1/2},\qquad N=|\mathcal W|.
\end{equation}
At criticality, frequencies were tracked at every integer overtone through $n=640$, whereas figure~\ref{fig:criticalresponses} uses discrete responses at $n=20,40,80,\allowbreak90,100,120,140,160,\allowbreak200,240,280,320,\allowbreak400,520,640$. The response samples were evaluated at the tracked roots by differentiating both exterior Frobenius recurrences, with increased precision and truncation and a changed matching point as checks. Panel~(a) connects all fifteen samples, with markers at six representative overtones; panel~(b) uses the thirteen with $n\ge80$.

For the retained critical expansions~\eqref{eq:criticalcompleteexpansions}, the full-response RMS on the six-point comparison set $n=20,40,80,160,320,640$ is $(0.031,0.027,0.24)\%$ for $(R_{\ell_1},R_{\ell_2},R_g)$, decreasing to $(0.00086,0.0013,0.027)\%$ on $n=160,320,640$; the isolated leading-coefficient discrepancy $|(R_{\rm num}-R_{\rm reg})/(A z^{-\rho})-1|$ at $n=640$ is instead $(0.32,0.44,0.25)\%$, with $(A,\rho)=(A_m,7/5)$ for both masses and $(A_g,11/10)$ for the gradient.
For the five timelike backgrounds, the complete fixed-coefficient response has RMS $0.0025\%$--$0.0051\%$ on the $80$ modes $n=161,\ldots,240$, improving the regular-only prediction by factors $142$--$169$, whereas $|A_{\rm eff}/A_c-1|$ from~\eqref{eq:effectivecoefficient} has RMS $0.59\%$--$0.69\%$ on the same window and discrepancy $0.54\%$--$0.64\%$ at $n=240$.
The improvement is smaller on lower and full timelike windows.
At the additional discrete response points, $|A_{\rm eff}/A_c-1|$ is $0.51\%$--$0.59\%$ at $n=280$ and $0.48\%$--$0.56\%$ at $n=320$; for each background, both values are smaller than its $n=240$ discrepancy. These checks extend the trend without enlarging the all-integer response window used for the RMS comparison.

\label{app:unresolved}
At a fixed zero-strength simple root, the response is linear in the inserted profile. Thus $R_{\ell_1}-R_{\ell_2}$ is the response to $H^{5/4}-H^{9/8}$: the common insertion $-\mathsf D$ cancels together with its geometric and endpoint corrections. Direct difference insertions in both exterior Frobenius recurrences at $n=80,320$ agree with separately computed responses. Increasing precision and truncation and changing the matching point alters the difference by less than $10^{-80}$ in relative terms, excluding subtraction loss as the source of the discrepancy at these tested points.

After subtracting the regular difference $[(B_1-B_2)\Log z+C_1-C_2]/z$, the two label terms retained in~\eqref{eq:criticalcompleteexpansions} have relative complex errors of about $15\%$ at $n=320$ and $9\%$ at $n=640$. A three-frequency filter that removes the logarithmic and constant regular terms remains more sensitive to later layers; its definition and comparison are included in the supplied numerical data. The retained-order analysis in appendix~\ref{app:coefficients} identifies no earlier missing contribution, but the present frequencies do not quantitatively validate these smaller label-dependent terms.

\bibliographystyle{JHEP}
\bibliography{references}
\end{document}